\documentclass[journal]{IEEEtran}
\usepackage{preambles}

\title{\textbf{Complex DNA Synthesis Sequences}}

\author{\IEEEauthorblockN{Boaz Moav, Eitan Yaakobi}
  \IEEEauthorblockA{Computer Science, Technion, Haifa, Israel \\
  Email: \{boazmoav, yaakobi\}@cs.technion.ac.il}
\and
\IEEEauthorblockN{Ryan Gabrys}
\IEEEauthorblockA{CalIT2, UCSD, San Diego, California, USA \\
Email: rgabrys@ucsd.edu}}

\author{Boaz Moav,~\IEEEmembership{Student,~IEEE,} Ryan Gabrys,~\IEEEmembership{Member,~IEEE,} Eitan Yaakobi,~\IEEEmembership{Senior Member,~IEEE}
\thanks{B. Moav and E. Yaakobi are with the Henry and Marilyn Taub Faculty of Computer Science, Technion - Israel Institute of Technology, Haifa 3200003, Israel 
(e-mail: \texttt{\{boazmoav,yaakobi\}@cs.technion.ac.il}). 
R. Gabrys is with CalIT2, University of California, San Diego, California, USA,
(e-mail: \texttt{rgabrys@ucsd.edu}).}
\thanks{%
The research was funded by the European Union (ERC, DNAStorage, 101045114 and EIC, DiDAX 101115134). Views and opinions expressed are, however, those of the authors only and do not necessarily reflect those of the European Union or the European Research Council Executive Agency. Neither the European Union nor the granting authority can be held responsible for them.
This work was also supported in part by NSF Grant CCF2212437.}
\thanks{%
This paper was presented in part at ISIT2025.}}
\begin{document}
\maketitle
\begin{abstract}
DNA-based storage offers unprecedented density and durability, but its scalability is fundamentally limited by the efficiency of parallel strand synthesis. Existing methods either allow unconstrained nucleotide additions to individual strands, such as enzymatic synthesis, or enforce identical additions across many strands, such as photolithographic synthesis. We introduce and analyze a hybrid synthesis framework that generalizes both approaches: in each cycle, a nucleotide is selected from a restricted subset and incorporated in parallel.
This model gives rise to a new notion of a \emph{complex synthesis sequence}. Building on this framework, we extend the information rate definition of Lenz et al. and analyze an analog of the deletion ball, defined and studied in this setting, deriving tight expressions for the maximal information rate and its asymptotic behavior. These results bridge the theoretical gap between constrained models and the idealized setting in which every nucleotide is always available.
For the case of known strands, we design a dynamic programming algorithm that computes an optimal complex synthesis sequence, highlighting structural similarities to the shortest common supersequence problem.
We also define a distinct two-dimensional array model with synthesis constraints over the rows, which extends previous synthesis models in the literature and captures new structural limitations in large-scale strand arrays. 
Additionally, we develop a dynamic programming algorithm for this problem as well.
Our results establish a new and comprehensive theoretical framework for constrained DNA, subsuming prior models and setting the stage for future advances in the field.
\end{abstract}

\section{Introduction}\label{sec:introduction}
DNA-based data storage has emerged as a promising alternative to conventional storage technologies, driven by advances in DNA synthesis and sequencing~\cite{Church2012,Goldman2013}. 
It offers orders-of-magnitude higher information density and long-term stability without power.
However, the synthesis stage remains expensive and time-consuming, motivating continued research.

A DNA storage system comprises three components:
a DNA synthesizer that produces the encoded strands, a storage medium for preserving them, and a DNA sequencer that reads and converts them back to digital form.
External encoding/decoding ensures reliable reconstruction even in the presence of errors.
While many parts of the pipeline have been optimized~\cite{Yazdi2017,Lenz2020_2,Rashtchian2017}, less attention has been given to the synthesis stage, which is the focus of this work.

Current DNA synthesis machines rely on parallel strand creation and typically follow two paradigms.
In the first method, the machine can add any of the four nucleotides in each cycle (e.g., photon-directed multiplexed enzymatic synthesis~\cite{Lee2020}).
In contrast, in the second method, nucleotides are added according to a predetermined synthesis sequence: in each cycle, a designated nucleotide is appended only to selected strands, implying that every synthesized strand must be a subsequence of this predefined (e.g., phosphoramidite chemistry~\cite{Kosuri2014} and photolithographic synthesis~\cite{Antkowiak2020}).

Prior work on the second paradigm analyzed the information density per cycle~\cite{Lenz2020, zrihan2024}, alternative alphabets~\cite{AbuSini2025}, and codes for efficient synthesis~\cite{SchouhamerImmink2024}, all with a single nucleotide option per cycle.
By contrast, the first paradigm yields a trivial upper bound on the information rate since all strands can extend every cycle. This creates a large theoretical gap, which we address by studying information rates when multiple nucleotides are available per cycle.

\begin{figure}
\centering
\includegraphics[width=\linewidth]{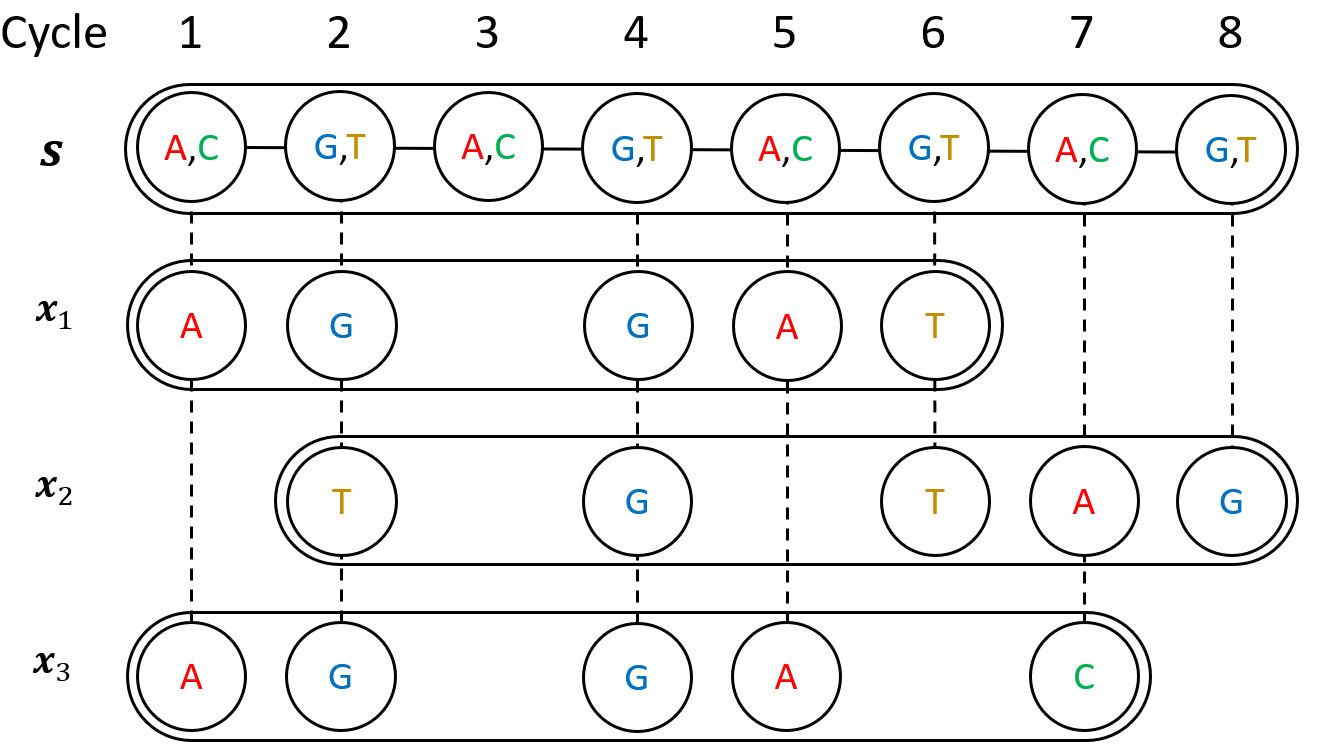}
\caption{Illustration of cycles of complex synthesis. Synthesizing the sequences $\bfx_1 = AGGAT$, $\bfx_2 = TGTAG$ and $\bfx_3 = AGGAC$ using $\bfs$, a periodic complex synthesis sequence of length $8$. 
Note that $\bfx_1,\bfx_2,\bfx_3$ are sub-instances of $\bfs$.
However, there is not a common supersequence of $\bfx_1,\bfx_2,\bfx_3$ which is an instance of $\bfs$ and $\bfs$ is not an shortest common complex supersequence of $\bfx_1,\bfx_2,\bfx_3$, since the third complex symbol of $\bfs$ is unused and can be removed.}
\label{fig:complex-synthesis}
\end{figure}

We introduce \emph{the complex DNA synthesis} model. 
In this model, a machine synthesizes over an alphabet of size $q$ (e.g., $q=4$ for the DNA alphabet). 
In each cycle, it has access to $w$ out of the $q$ nucleotides and may append one of these $w$ nucleotides to any strand, with at most one nucleotide per strand per cycle (see \Cref{fig:complex-synthesis}).
Our framework generalizes the $w=1$ case studied by Lenz et al.~\cite{Lenz2020} and the trivial $w=q$ case, where all nucleotides are always available~\cite{Lee2020}.
Restrictions on subsets of nucleotides may arise in practice in the future.

Within this framework, we study the information rate following the definition of~\cite{Lenz2020}, which describes the asymptotic maximum number of information bits stored per synthesis cycle.
We extend the notions of subsequences and supersequences to relate standard sequences to complex synthesis sequences and identify sequences that maximize the rate. 

In this work, we improve our results from~\cite{Moav2025} by deriving closed-form expressions for the maximal rate as a function of $q$ and $w$. These expressions are detailed in \Cref{theorem:f(q-w)-using-singularity} and \Cref{theorem:max-information-rate}, and numerical values are summarized in \Cref{table:summary-f-q-w-info-rate}.
As shown there, even for the DNA setting ($q=4$), the information rate increases substantially for $w=2$ or $w=3$ (for instance, from $0.947$ to $1.964$ when $w$ increases from $1$ to $3$), supporting further investigation into practical synthesis methods inspired by this theoretical framework.
For known strands, we additionally provide a dynamic programming algorithm that computes an optimal complex synthesis sequence. Our results are consistent with the $w=1$ setting of~\cite{Lenz2020}.
To the best of our knowledge, this is the first analysis of the information rate under multiple nucleotide options per cycle.

\renewcommand{\arraystretch}{1.4}
\begin{table*}[t]
\centering
\caption{Values of the tuple $(f(q,w),\alpha)$, where $\alpha$ is the value such that $f(q,w)=f(q,w,\alpha)$, rounded to 3 decimal places.}
\begin{tabular}{|| c | c | c | c | c | c | c | c | c ||}
\hline
\diagbox[width=0.7cm, height=0.6cm]{\hspace*{-3pt}$q$}{$w$\hspace*{-3pt}} & $1$ & $2$ & $3$ & $4$ & $8$ & $16$ & $64$ & $256$
\\ [0.5ex] 
\hline\hline
$2$ & $(0.694,0.724)$ & $1$ & --- & --- & --- & --- & --- & ---
\\
\hline
$3$ & $(0.879,0.618)$ & $(1.272,0.854)$ & $1.585$ & --- & --- & --- & --- & ---
\\
\hline
$4$ & $(0.947,0.566)$ & $(1.450,0.789)$ & $(1.724,0.916)$ & $2$ & --- & --- & --- & ---
\\
\hline
$8$ & $(0.997,0.507)$ & $(1.573,0.687)$ & $(1.964,0.791)$ & $(2.272,0.854)$ & $3$ & --- & --- & ---
\\
\hline
$16$ & $(1.00,0.500)$ & $(1.585,0.667)$ & $(1.999,0.752)$ & $(2.320,0.804)$ & $(3.154,0.908)$ & $4$ & --- & ---
\\
\hline
$64$ & $(1.000,0.500)$ & $(1.585,0.667)$ & $(2.000,0.750)$ & $(2.322,0.800)$ & $(3.170,0.889)$ & $(4.087,0.941)$ & $6$ & ---
\\
\hline
$256$ & $(1.000,0.500)$ & $(1.585,0.667)$ & $(2.000,0.750)$ & $(2.322,0.800)$ & $(3.170,0.889)$ & $(4.087,0.941)$ & $(6.022,0.985)$ & $8$
\\
\hline
\end{tabular}
\label{table:summary-f-q-w-info-rate}
\end{table*}

Recent experimental advances in large-scale, light-directed, and enzymatic DNA synthesis have revealed that spatial control across dense arrays is inherently limited by optical and chemical coupling between neighboring sites. The works of \cite{Sack2013, Antkowiak2020} identified scattering, diffraction, and optical-flare effects as the dominant sources of sequence errors in earlier light-directed systems, motivating their redesign of the photochemical reaction cell, in which two substrates must be positioned approximately 60-100 $\mu$m apart to minimize reflections. These findings demonstrate that maintaining synthesis fidelity depends on restricting simultaneous activations within shared optical or chemical regions of the array.

Motivated by these experimentally observed constraints, we model an abstract DNA synthesis process on an $m \times n$ array, in which at most $\rho$ strands per row advance during each cycle, while multiple columns may proceed concurrently. This abstraction captures the effective resource coupling observed in the experimental system of Sack et al.~\cite{Sack2013} and provides a theoretical framework for analyzing the efficiency limits of spatially constrained parallel synthesis.

Recall from Figure~\ref{fig:complex-synthesis} that the primary challenge involved in the complex synthesis process was determining which of the $w$ possible nucleotides to assign to a collection of, say, $k$ strands, where we are allowed to assign any of the $w$ nucleotides to as many of the $k$ strands as we would like.
In \Cref{sec:2d-array}, we initiate the study of the setting where we can assign each of the $w$ nucleotides to at most a certain number of programmed strands.
As a starting point, in this paper, we begin with the setting where $k=2$, $w=1$, and each nucleotide from the synthesis strand can be assigned to at most a single strand.

Even in this simple setting, the problem does not have a straightforward answer, and many of the techniques developed previously appear incompatible under the new framework. For this setting, we design a dynamic-programming algorithm and outline open questions that may be addressed in future research.

The distinction between the fully parallel setting and the array model lies in how much parallelism is permitted per cycle. 
In the standard and complex synthesis settings, strands do not interfere with one another: in each cycle, any strand whose next base lies in the available set (of size $w$) may advance. 
By contrast, the array model enforces row-level coupling. That is, at most one strand per row may advance in a given cycle.

For a concrete comparison, let $w=1$, the periodic synthesis sequence $(ACGTACGT\ldots)$, and target strands $ACG$ and $CGT$.
Then, in the fully parallel model, both are completed in four cycles. 
Under the array constraint, if the two strands share a row, only one may advance per cycle; with the fixed $ACGT$ phase, the shortest completion takes seven cycles (e.g., synthesize $CGT$ using cycles 2-4, and $ACG$ using cycles 1,6, and 7), because in the second cycle one cannot synthesize $C$ in both strands.
This added coupling fundamentally changes the optimization compared with the fully parallel case.

The rest of the paper is organized as follows:
\Cref{sec:definitions-and-problems} introduces preliminaries, definitions, problem statements, and related work; \Cref{sec:information-rate} develops information rate results for the complex DNA synthesis; \Cref{sec:SCCS} presents algorithms for determining an optimal synthesis sequence; and \Cref{sec:2d-array} provides initial results for the $m \times n$ array model.

\section{Definitions and Problem Statement}\label{sec:definitions-and-problems}
\subsection{Definitions and Preliminaries}\label{subsec:definitions}

Let $\Sigma_{q}$ be an alphabet of size $q$. We first define a complex symbol and a complex sequence.

\begin{definition}\label{def:complex-symbol}
Let $q$ and $w \leq q$ be positive integers. Denote 
$$\Psi_{q,w} \coloneqq \binom{\Sigma_q}{w} = \{\bfpsi \subseteq \Sigma_q \colon |\bfpsi| = \!w \},$$
as the set of all $\binom{q}{w}$ sets of distinct $w$ elements from $\Sigma_{q}$.
We define a \emph{complex symbol} as 
$$
    \bfpsi \!=\! \{ \sigma^{(1)}, \ldots, \sigma^{(w)} \} \!\in\! \Psi_{q,w},
$$
where $\sigma^{(1)}, \ldots, \sigma^{(w)} \!\in\! \Sigma_q$, and a \emph{complex sequence} as 
$$
    \bfs = (\bfpsi_{1}, \bfpsi_{2}, \ldots, \bfpsi_{n}) \in \Psi_{q,w}^{n}.
$$
\end{definition}

Note that $\Psi_{q,1}$ is isomorphic to $\Sigma_q$.
Also, throughout this paper, we denote arbitrary characters in $\Sigma_q$ as $\sigma$, and arbitrary complex symbols in $\Psi_{q,w}$ as $\bfpsi$. For integers $n>0$ and $0 \leq t \leq n$, let $\bfD(\bfx,t)$ be the $t$-deletion ball of $\bfx\in\Sigma_{q}^{n}$, and similarly, let $\bfD(\bfs,t)$ be the $t$-deletion ball of $\bfs\in\Psi_{q,w}^{n}$, i.e., deletions of complex symbols.
An example illustrating \Cref{def:complex-symbol} is given below.
\begin{example}\label{example:complex-sym-and-complex-seq}
Let $\Sigma_4\!=\!\{A,C,G,T\}$, and the complex symbols,
\begin{align*}
    \Psi_{4,2}=\{ &\{A, C\}, \{A, G\}, \{A, T\}, \\
    &\{C, G\}, \{C, T\}, \{G, T\} \}.
\end{align*}
Two complex sequences of length $4$ over $\Psi_{4,2}$ are,
\begin{align*}
    \bfs_1&=( \{A, C\}, \{A, G\}, \{A, T\}, \{A, T\} ), \\
    \bfs_2&=( \{A, C\}, \{G, T\}, \{A, C\}, \{G, T\} ).
\end{align*}
The single-deletion ball of $\bfs_1$ consists of 
\begin{dmath*}
    \bfD(\bfs_1,1) \!=\! \{ ( \{A, G\},\allowbreak \{A, T\},\allowbreak \{A, T\} ),\allowbreak ( \{A, C\},\allowbreak \{A, T\},\allowbreak \{A, T\} ),\allowbreak ( \{A, C\},\allowbreak \{A, G\},\allowbreak \{A, T\} ) \}.
\end{dmath*}
\end{example}

Next, we generalize the notion of a subsequence so that we can say that $\bfx=(x_1, x_2, \ldots, x_n)\in\Sigma_{q}^{n}$ is ``a subsequence'' of $\bfs \in \Psi_{q,w}^{*}$. 

\begin{definition}\label{def:instance-of-sequence}
We say that $\bfx\in\Sigma_{q}^{n}$ is an \emph{instance} of $\bfs\in\Psi_{q,w}^{n}$ if $x_i \in \bfpsi_i$, for all $1 \leq i \leq n$.
Let $\cI(\bfs)$ denote the set of all instances of $\bfs$. Moreover, $\bfy\in\Sigma_q^n$ is a \emph{sub-instance} of $\bfs\in\Psi_{q,w}^{*}$ if $\bfy$ is a subsequence of some instance $\bfx$ of $\bfs$.
\end{definition}

Note that $|\cI(\bfs)|=w^n$ for all $\bfs\in\Psi_{q,w}^{n}$.
An example for \Cref{def:instance-of-sequence} is presented next.

\begin{example}\label{example:sub-instance}
    Let $\bfs_1$ be as defined in \Cref{example:complex-sym-and-complex-seq}. Two instances are 
    \begin{align*}
        \bfx_1=( A, A, A, T ),\quad
        \bfx_2=( A, G, A, T).
    \end{align*}
    Two example sub-instances of $\bfs_1$ are 
    \begin{align*}
        \bfy_1=( A, A, A ),\quad
        \bfy_2=( A, A, T ).
    \end{align*}
\end{example}

Next, we define an analog of a $t$-deletion-error ball for complex sequences.

\begin{definition}\label{def:sub-instance-ball}
    Let $\bfs\in\Psi_{q,w}^{n}$. Denote the \emph{$t$-sub-instance ball},
    $$\cD_{q,w}(\bfs,t) = \bigcup_{\bfx\in\cI(\bfs)} \bfD(\bfx,t)\,.$$
\end{definition}

For shortening, if $q$ and $w$ are clear from the context, we will write $\cD_{q,w}(\bfs,t)$ as $\cD(\bfs,t)$. 
Moreover, note that $\cD(\bfs,t)$ contains all the sub-instances of $\bfs$ of length $n-t$. According to Definition~\ref{def:sub-instance-ball}, the $t$-sub-instance ball $\cD(\bfs,t)$ is the set of sequences obtainable by first considering the set of all instances of $\bfs$ and then deleting $t$ symbols from every sequence in the resulting set. 

In \Cref{claim:sub-instance-ball-equivalence}, we show that $\cD(\bfs,t)$ can also be obtained by reversing the order of operations and first deleting $t$ complex symbols from $\bfs$, and then considering all sub-instances of the resulting set of sequences.


\begin{claim}\label[Claim]{claim:sub-instance-ball-equivalence} Let $\bfs\in\Psi_{q,w}^{n}$. Then,
    \begin{align*}
        \cD(\bfs,t) = \bigcup_{\bfx\in\cI(\bfs)} \bfD(\bfx,t) = \bigcup_{\bfx\in\bfD(\bfs,t)}\cI(\bfx)\,.
    \end{align*}
\end{claim}
\begin{IEEEproof}
The proof is by double inclusion.
First, if $\bfy \in \cD(\bfs,t)$, then there exists $\bfx\in\cI(\bfs)$, such that if $i_1, i_2, \ldots, i_t$ entries of $\bfx$ are deleted, it results in $\bfy$.
Thus, deleting the same indices from $\bfs$, yields $\bfs^{\prime}$, and clearly $\bfy$ is a sub-instance of $\bfs^{\prime}$.
Conversely, if $\bfw \in \bigcup_{\bfx\in\bfD(\bfs,t)}\cI(\bfx)$, there exists $\bfs^{\prime}$, such that if $i_1, i_2, \ldots, i_t$ entries of $\bfs$ are deleted, it results in $\bfs^{\prime}$ and $\bfw$ is a sub-instance of $\bfs^{\prime}$.
Hence, if one adds an arbitrary character from the original characters in the complex symbol in the positions $i_1, i_2, \ldots, i_t$ of $\bfs$, in their exact position, we obtain a sequence $\bfz$, such that $\bfw\in\bfD(\bfz,t)$ and $\bfz\in\cI(\bfs)$.
\end{IEEEproof}

Note the distinction in \Cref{claim:sub-instance-ball-equivalence} between $\cD(\bfs,t)$ (scripted $\cD$), which is a set of sub-instances, i.e., sequences over $\Sigma_q$, and $\bfD(\bfs,t)$ (non-scripted $\bfD$), which is a set of complex sequences, i.e., sequences over $\Psi_{q,w}$.

Although the number of instances is the same for any $\bfs$, the number of sub-instances strongly depends on the structure of the complex sequence itself.
This variability in sub-instance distribution parallels the behavior observed in the classical setting of deletions over non-complex strings.
Finally, while this appears to be an abuse of notation, note that $\cD(\bfs,t)=\bfD(\bfs,t)$ when $w=1$, justifying our choice.
We next generalize the notion of a run in a sequence, which we will use to analyze the single-sub-instance balls in~\Cref{subsec:sub-instance-ball}.

\begin{definition}\label{def:intersection-index}
    Let $\bfs = (\bfpsi_{1}, \bfpsi_{2}, \ldots, \bfpsi_{n}) \in \Psi_{q,w}^{n}$. Let the \emph{complex runs} of $\bfs$ be the integer sequence $r({\bfs})$ of length $n$, where $(r({\bfs}))_{1}=w$ and for every $2 \leq j \leq n$, $(r({\bfs}))_{j}=|\bfpsi_{j} \setminus \bfpsi_{j-1}|$.
    Moreover, denote the \emph{type-$k$ complex runs} by $r_{k}(\bfs)$ for any $0 \leq k \leq w$, as the number of entries $(r(\bfs))_{j}$ with the value $k$.
\end{definition}

\begin{example}\label{example:intersection-index}
Let $q=5$, $w=3$, $\Sigma_5=\{A,C,G,T,M\}$,
\begin{dmath*}
    \bfs\!=\!( \{A, C, G\},\!\allowbreak \{A, T, M\},\!\allowbreak \{C, G, M\},\!\allowbreak \{C, G, T\},\allowbreak \{G, T, M\},\allowbreak \{G, T, M\} ).
\end{dmath*}
Then, it holds that $r(\bfs) = (3,2,2,1,1,0)$, $r_0(\bfs) = r_3(\bfs) = 1$, and, $r_1(\bfs) = r_2(\bfs) = 2$.
\end{example}

\subsection{Problem Statement}\label{subsec:problems}
Extending the model introduced in~\cite{Lenz2020}, consider a system where $k$ DNA strands $\bfx_1, \bfx_2, \ldots, \bfx_k \in \Sigma_q^{*}$ are synthesized in parallel.
In our model, synthesis proceeds by choosing a fixed complex synthesis sequence $\bfs = (\bfpsi_{1}, \bfpsi_{2}, \ldots) \in \Psi_{q,w}^{*}$.
In each cycle, for each DNA strand $\bfx_j$, we may either append one nucleotide from $\bfpsi_i$ to the strand $\bfx_j$ or do nothing.
A strand $\bfx$ can be synthesized by $\bfs$ if and only if $\bfx$ is a sub-instance of $\bfs$.

For sequences $\bfs \in \Psi_{q,w}^{*}$ and $\bfx \in \Sigma_q^{*}$, define $\bfs_{i:j} \coloneqq (\bfpsi_i, \bfpsi_{i+1},\ldots,\bfpsi_j)$ and $\bfx_{i:j} \coloneqq (\sigma_i, \sigma_{i+1},\ldots,\sigma_j)$ as contiguous subsequences. 
Let $N^{n}(\bfs,\tau)$ denote the number of distinct sequences over $\Sigma_{q}^{n}$ that can be synthesized by $\bfs_{1:\tau}$; let $N(\bfs,\tau)$ denote the number over $\Sigma_{q}^{*}$. 
The following lemma is immediate.

\begin{lemma}\label{lemma:number-of-sub-instances}
Let $\bfs\in\Psi_{q,w}^{*}$, and let $\tau$ be a positive integer. Then,
\begin{align*}
    N^{n}(\bfs, \tau) &= |\cD(\bfs_{1:\tau}, \tau-n)| \\
    N(\bfs, \tau) &= \sum_{i=0}^{\tau}|\cD(\bfs_{1:\tau},i)|
\end{align*}
\end{lemma}
\begin{IEEEproof}
    Every $\bfy \in N^{n}(\bfs, \tau)$ is, by definition, a length-$n$ sub-instance of $\bfs_{1:\tau}$, hence $\bfy \in \cD(\bfs, \tau-n)$.
    The converse is analogous. 
    Summing over $n$ yields the second identity.
\end{IEEEproof}

\begin{definition}\label{def:information-rate}
Let $\bfs\in\Psi_{q,w}^{*}$. We define the \emph{information rates} as follows.
\begin{align*}
    \cR_{q,w}(\bfs,\alpha) &= \limsup_{t\to\infty}\frac{\log_2\left(N^{\lfloor\alpha t\rfloor}(\bfs,t)\right)}{t}\,, \\
    \cR_{q,w}(\bfs) &= \limsup_{t\to\infty}\frac{\log_2\left(N(\bfs,t)\right)}{t}\,.
\end{align*}
Then, denote
$$
    f(q,w,\alpha) \coloneqq \max_{\bfs\in\Psi_{q,w}^{*}}\cR_{q,w}(\bfs, \alpha).
$$
And similarly, 
$$
    f(q,w) \coloneqq \max_{\bfs\in\Psi_{q,w}^{*}}\cR_{q,w}(\bfs).
$$
\end{definition}
We now state the problems we address.
\begin{problem}\label[Problem]{problem:optimal-information-rate}
    Compute $f(q,w,\alpha)$ and $f(q,w)$, and find semi-infinite sequences $\bfs\in\Psi_{q,w}^{*}$ that attain these values.
\end{problem}

\Cref{sec:information-rate} presents our results for \Cref{problem:optimal-information-rate}.
\Cref{def:information-rate} coincides with~\cite{Lenz2020}; \cite[Theorem 3]{Lenz2020} solves $w=1$. 
For $w=q$, $f(q,q) = \log_2(q)$, because all strands advance every cycle; this serves as a trivial upper bound.
\Cref{problem:optimal-information-rate} fixes the complex synthesis sequence.
We next consider the case where the sequence is chosen based on the given set of $k$ strands.
\begin{definition}\label{def:SCCS}
Let $\cX = \{ \bfx_1, \bfx_2, \ldots, \bfx_k \} \subseteq \Sigma_{q}^{*}$, and denote $\text{SCCS}(\cX)\!\in\!\Psi_{q,w}^{*}$ as a shortest common complex supersequence of the sequences of $\cX$.
That is, a shortest complex sequence $\bfs$ such that $\bfx_j$ ($1 \leq j \leq k$) is a sub-instance of $\bfs$.
\end{definition}
An SCCS is a valid complex synthesis sequence and, by definition, minimizes the number of cycles required to synthesize the $k$ strands.
In \Cref{sec:SCCS}, we find an SCCS for a set of sequences using a dynamic programming algorithm.
\begin{problem}\label[Problem]{problem:shortest-synthesis-sequence}
    Given a set of sequences $\cX$, compute an $\text{SCCS}(\cX)$.
\end{problem}

\subsection{Previous Work}\label{subsec:previous-work}
The problem of efficient synthesis was first introduced in~\cite{Lenz2020}, where the goal was to determine the maximum number of nucleotides that can be encoded across multiple strands given $t$ synthesis cycles.
In particular, \cite{Lenz2020} solved \Cref{problem:optimal-information-rate} for the case $w=1$.
In that work, the authors proved that for $0 \leq \alpha \leq 1$, the sequence that maximizes the rate is the periodic sequence $\bfp_q$, i.e., 
$$\bfp_q=(\sigma_1, \sigma_2, \ldots, \sigma_q, \sigma_1, \sigma_2, \ldots, \sigma_q, \ldots).$$

Moreover, the authors derived the exact values of $f(q,1)$ and $f(q,1,\alpha)$, as stated in~\cite[Theorem 3]{Lenz2020} and~\cite[Proposition 6.7]{Lenz2021}. 
Their main results can be summarized as follows:
\begin{theorem}[\hspace{1sp}\cite{Lenz2021}]\label{theorem:f-for-w=1}
Let $q$ be an integer and $0 \leq \alpha < 1$. Then,
$
    f(q,1) = -\log_2(x_q)\,,
$
where $x_q$ is the unique positive solution to $\sum_{i=1}^{q} x^{i} = 1$. Also,     
\footnotesize \begin{align*}
    f(q,1,\alpha) &=\begin{cases}
    \alpha \log_2(q) &\, \alpha < \frac{2}{q+1}\,, \\
    \alpha\log_2\left(\alpha\sum_{i=1}^{q}i\cdot(x_{q,\alpha})^{i-\frac{1}{\alpha}}\right) &\, \frac{2}{q+1} < \alpha < 1,
    \end{cases}
\end{align*}
\normalsize where $x_{q,\alpha}$ is the unique solution to $\sum_{i=1}^{q} (1-\alpha i)x^{i}=0$, on the interval $0<x<1$.
\end{theorem}

In addition, \cite{Lenz2020} presented an explicit code construction for the periodic synthesis sequence $\bfp_q$.
They showed that any strand of length $n-1$ can be encoded into a sequence of length $n$, synthesizable in $\left\lfloor\frac{q+1}{2}n\right\rfloor$ cycles. 
Subsequent works expanded on these results.
In \cite{zrihan2024}, new codes were proposed that achieve a high information rate, approaching the capacity. 
The authors used the same periodic synthesis sequence of length $n$, and introduced coding schemes for any $0 < \rho < 1$, to synthesize strands of length $\rho n$.
In~\cite{SchouhamerImmink2024}, the authors generalized the synthesis codes by introducing a family of low-rate codes, and in~\cite{SchouhamerImmink2024_2} the same authors analyzed the redundancy and information rate of such constructions for asymptotically large codeword lengths.
In parallel, \cite{AbuSini2025} considered an alternative setting in which shortmers were used as synthesis symbols. 

Also, in~\cite{Makarychev2022} the authors introduced the problem of synthesis in batches, i.e., to partition the sequences to be synthesized into batches of equal size, and synthesize each batch separately, to minimize the number of synthesis cycles.
The authors presented bounds on the number of cycles that are required to synthesize all the batches, using an optimal partition of the sequences beforehand.
In~\cite{Singh2023}, the authors gave a novel approach to optimize the bounds of~\cite{Makarychev2022} with a polynomial-time concurrent zero-knowledge proof protocol.

In~\cite{Chrisnata2023}, a single-indel-correcting code was presented, which is also limited by the number of synthesis cycles needed to synthesize each codeword, i.e., each strand. In other words, the authors constructed an explicit VT-code, which has only codewords that can be synthesized in $T$ cycles for a fixed $T$. Additionally, in~\cite{Nguyen2024}, the authors improved the code of~\cite{Chrisnata2023} and also presented another code construction, which satisfies the run-length-limited constraint as well. Moreover, in~\cite{Lu2024}, the authors presented codes and explicit constructions for correcting errors that are the result of a synthesis defect, i.e., when the synthesis machine does not add the required nucleotide in a given cycle to any of the strands.
Also, in~\cite{liu2025constrainederrorcorrectingcodesefficient} the authors proposed a code that can be synthesized in $T$ synthesis cycles while maintaining the $\ell$-RLL and $\epsilon$-balanced constraints. They provided an explicit algorithm for encoding and decoding sequences of length $n$ in a runtime complexity of $O(n^4\ell^2)$, and a capacity-achieving redundancy. The authors also provided an extension to handle indel errors with an additional redundancy.

These prior works fully characterize the case $w=1$; in the next section, we broaden the scope to $w>1$ and establish the information-theoretic limits of the complex synthesis model.

\section{Information Rate}\label{sec:information-rate}

In this section, we address \Cref{problem:optimal-information-rate}. In \Cref{subsec:sub-instance-ball}, we study the size of the sub-instance ball.
In \Cref{subsec:information-rate-analysis}, we use these results to compute the maximal information rate for complex sequences.

\subsection{The Size of the Sub-instance Ball}\label{subsec:sub-instance-ball}

In this subsection, we study $\cD(\bfs,t)$, the $t$-sub-instance ball of a complex synthesis sequence $\bfs = (\bfpsi_{1}, \bfpsi_{2}, \ldots, \bfpsi_{n}) \in \Psi_{q,w}^{n}$. We generalize definitions and results from~\cite{Lenz2020} to handle complex sequences.
For $\sigma\in\Sigma_q$ and $\bfA\subseteq\Sigma_q^{*}$, define $\sigma \circ \bfA \coloneqq \{\bfx=(\sigma,y_1,\ldots,y_n) \colon \allowbreak (y_1,\ldots,y_n)\in\bfA\}$.

\begin{definition}
    Let $\Sigma_{\bfs,t} = \cup_{i=1}^{t+1}\bfpsi_i$, i.e., the set of all symbols from $\Sigma_q$ appearing in the first $t+1$ complex symbols of $\bfs$. 
    For $\sigma \in \Sigma_{\bfs,t}$, let $j_{\sigma}$ denote the index of the first complex symbol $\bfpsi_{j_\sigma}$ in $\bfs$ that contains $\sigma$.
\end{definition}

\begin{lemma}\label{lemma:partitioning-the-sub-instance-ball}
    Let $0<n,t\in\mathbb{N}$ and $\bfs\in\Psi_{q,w}^{n}$.
    Then,
    $$\cD(\bfs,t) = \bigcup_{\sigma\in\Sigma_{\bfs,t}} \sigma \circ \cD(\bfs_{j_{\sigma}+1:n},t-j_{\sigma}+1).$$
    Consequently,
    $$|\cD(\bfs,t)| \,\,= \sum_{\sigma\in\Sigma_{\bfs,t}} |\cD(\bfs_{j_{\sigma}+1:n},t-j_{\sigma}+1)|.$$
\end{lemma}
\begin{IEEEproof}
This result follows immediately from the same arguments as~\cite[Lemma 1]{Hirschberg1999}.
\end{IEEEproof}

It is well known (see, e.g.,~\cite{Levenshtein1965}) that the size of the single-deletion ball of $\bfx\in\Sigma_q$ equals the number of runs in $\bfx$.
The following theorem generalizes this result to complex sequences.
Recall from \Cref{def:intersection-index} that $r(\bfs)$ denotes the vector of complex runs of $\bfs$, and $r_k(\bfs)$ denotes the number of type-$k$ complex runs, i.e., the entries of $r(\bfs)$ equal $k$).
\begin{theorem}\label{theorem:single-sub-instance-ball}
     Let $n>1$ and $\bfs\in\Psi_{q,w}^{n}$. Then, 
     \begin{align*}
         |\cD(\bfs,1)| = w^{n-2} \, \sum_{k=1}^{w}k \, r_k(\bfs)\,.
     \end{align*}
\end{theorem}
\begin{IEEEproof}
    Since $t=1$, \Cref{lemma:partitioning-the-sub-instance-ball} implies the following recursion:
    \begin{align}
        |\cD(\bfs,1)| &= |\bfpsi_1| \cdot |\cD(\bfs_{2:n},1)| + |\bfpsi_2 \setminus \bfpsi_1| \cdot |\cD(\bfs_{3:n},0)| \nonumber \\
        &= w \cdot |\cD(\bfs_{2:n},1)| + (r(\bfs))_2 \cdot | \cI(\bfs_{3:n}) | \nonumber \\
        &= w \cdot |\cD(\bfs_{2:n},1)| + (r(\bfs))_2 \cdot w^{n-2}\,. \label{eq:rec-expr-for-1-si-ball}
    \end{align}
    Thus, we provide a proof using induction on $n$.
    For $n=2$, $\bfs=(\bfpsi_1,\bfpsi_2)$. Then by \Cref{lemma:partitioning-the-sub-instance-ball}, $|\cD(\bfs,1)|= |\bfpsi_1| + |\bfpsi_2 \setminus \bfpsi_1| = w + (r(\bfs))_2$.
    Since $w^{n-2}=1$, and $r_k(\bfs)=0$ except possibly for $k=w$ and $k=(r(s))_2$, the expression $w^{n-2}\sum_{k=1}^{w} k r_k(\bfs)$ equals $w+(r(s))_2$, as required.

    We proceed by induction on $n$. 
    Using \eqref{eq:rec-expr-for-1-si-ball} and the inductive hypothesis,
    \begin{align*}
        |\cD(\bfs,1)| &= w \left( w^{n-3} \cdot \sum_{k=1}^{w}k\cdot r_k(\bfs_{2:n}) \right) + (r(\bfs))_2 \cdot w^{n-2} \\
        &= w^{n-2} \left(\sum_{k=1}^{w}k r_k(\bfs_{2:n}) + (r(\bfs))_2 \right) .
    \end{align*}
    Since $\sum_{k=1}^{w}k\cdot r_k(\bfs) - \sum_{k=1}^{w}k\cdot r_k(\bfs_{2:n}) = (r(\bfs))_2$, the claim follows, which completes the proof.    
\end{IEEEproof}

Note that for $w=1$, the value of $r_{1}(\bfx)$ for a (traditional) sequence $\bfx\in\Sigma_q^{*}$ is exactly the number of runs in $\bfx$, and thus \Cref{theorem:single-sub-instance-ball} coincides with the result for the single-deletion ball when $w=1$.

\begin{example}\label{example:single-sub-instance-ball}
    Using the notation from \Cref{example:intersection-index}, we get that
    \begin{align*}
         |\cD(\bfs,1)| 
         = 3^{4} \cdot ( r_1(\bfs) + 2\cdot r_2(\bfs) + 3\cdot r_3(\bfs))
         = 729\,.
     \end{align*}
\end{example}
\begin{claim}
    Let $n$, $q$ and $w\leq q$ be positive integers, and let $\bfs\in\Psi_{q,w}^{n}$. Then, 
    \begin{enumerate}
        \item $w^{n-1} \leq |\cD(\bfs,1)|$, and the constant sequences of length $n$, $\bfs^{n}_{\bfpsi}=(\bfpsi,\bfpsi,\ldots,\bfpsi)\in\Psi_{q,w}^{n}$, attain the minimum of the single-sub-instance ball.
        \item If $w\leq \left\lfloor\frac{q}{2}\right\rfloor$, then $|\cD(\bfs,1)| \leq w^{n-1} \cdot n$, and the sequences $\bfs^{(1)}=(\bfpsi_1, \bfpsi_2, \ldots) \in \Psi_{q,w}^{n}$, such that $\bfpsi_{i}\cap\bfpsi_{i+1} = \emptyset$ for all $1 \leq i < n$, are the only sequences that attain the maximum of the single-sub-instance ball.
        \item If $w > \left\lfloor\frac{q}{2}\right\rfloor$, then $|\cD(\bfs,1)| \leq w^{n-1} + w^{n-2}(n-1)(q-w)$, and the sequences $\bfs^{(2)}=(\bfpsi_1, \bfpsi_2, \ldots) \in \Psi_{q,w}^{n}$, such that $|\bfpsi_{i}\cap\bfpsi_{i+1}| = 2w-q$ for all $1 \leq i < n$, are the only sequences that attain the maximum of the single-sub-instance ball.
    \end{enumerate}
\end{claim}
\begin{IEEEproof}
    Since 
    $$|\cD(\bfs,1)| = w^{n-2} \cdot \sum_{k=1}^{w}k\cdot r_k(\bfs),$$
    and if  $w\leq \left\lfloor\frac{q}{2}\right\rfloor$, the maximum is obtained if $r(\bfs)=(w,w,\ldots,w)$ for, hence $r_w(\bfs)=n$ and $r_k(\bfs)=0$ for all $0 < k < w$, i.e., the maximal value in each entry of $r(\bfs)$.
    Then, we note that $\bfs^{(1)}$ is the only structure of the complex sequences for which it holds that $r(\bfs^{(1)})=(w,w,\ldots,w)$.

    Otherwise, if $w > \left\lfloor\frac{q}{2}\right\rfloor$,  the maximum is obtained when $r(\bfs)=(w,q-w,\ldots,q-w)$, since one cannot have two adjacent symbols with an empty intersection, and $q-w$ is the minimal intersection size.
    
    It is obtained by choosing the symbol $\bfpsi_i$ to contain the $q-w$ characters that did not appear in $\bfpsi_{i-1}$, and thus leave $w-(q-w)=2w-q$ as $|\bfpsi_{i}\cap\bfpsi_{i+1}|$.
    Then, $\bfs^{(2)}$ is the only structure of the complex sequences for which it holds that $r(\bfs^{(2)})=(w,q-w,\ldots,q-w)$.

    The minimum is obtained if $r(\bfs^{n}_{\bfpsi})=(w,0,0,\ldots,0)$, because the first entry is always $w$, and thus $r_w(\bfs^{n}_{\bfpsi})=1$ and $r_k(\bfs^{n}_{\bfpsi})=0$ for all $0 < k < n$.
    Sequences of the form $\bfs^{n}_{\bfpsi}$ are the only complex sequence for which it holds that $r(\bfs^{n}_{\bfpsi})=(w,0,0,\ldots,0)$.
\end{IEEEproof}

The rest of this subsection is based on generalizations of results from~\cite{Hirschberg1999} that hold for complex sequences.
Let $\{\sigma_0,\sigma_1,\ldots,\sigma_{q-1}\}$ be some ordering of $\Sigma_q$.
Then, denote the \emph{periodic complex sequence} as $\bfc_n=(c_1,c_2,\ldots,c_n)\in\Psi_{q,w}^{n}$, where for every $1 \leq i \leq n$,
\begin{align*}
    c_i=\{ &\sigma_{(w(i-1))\Mod{q}},\allowbreak \sigma_{(1+w(i-1))\Mod{q}}, \ldots, \\
    &\sigma_{(wi-1)\Mod{q}}\}.
\end{align*}

\begin{example}\label{example:periodic-complex-sequence}
    Let $q=6, w=4, n=4$. Then, 
    \begin{gather}
        \Sigma_6=\{\sigma_0\!=0,\sigma_1\!=1,\sigma_2\!=2,\sigma_3\!=3,\sigma_4\!=4,\sigma_5\!=5\}\,, \nonumber\\
        \bfc_n = (\{0,1,2,3\}, \{4,5,0,1\}, \{2,3,4,5\}, \{0,1,2,3\}) .\nonumber
    \end{gather}
\end{example}

Our goal is to show that for every $\bfs\in\Psi_{q,w}^{n}$ and any $t$, $|\cD(\bfs,t)| \leq |\cD(\bfc_n,t)|$.
Let $m=\left\lfloor\frac{q}{w}\right\rfloor$.
First, a lemma is given to obtain a recursive formula of the $t$-sub-instance ball of the periodic complex sequence.

\begin{lemma}\label{lemma:recursive-formula-of-t-sub-instance-ball-for-cyclic}
    Let $1 \leq t \leq n-1$. Then,
    \begin{align*}
        |\cD(\bfc_n,t)| &= w \cdot \sum_{i=1}^{m}|\cD(\bfc_{n-i},t-i+1)| \\ 
        &+ (q-w\cdot m) \cdot |\cD(\bfc_{n-m-1},t-m)|.
    \end{align*}
    In particular, if $w \mid q$, it holds that,
    $$|\cD(\bfc_n,t)| = w \cdot \sum_{i=1}^{m}|\cD(\bfc_{n-i},t-i+1)|.$$
\end{lemma}
\begin{IEEEproof}
    Using \Cref{lemma:partitioning-the-sub-instance-ball}, it holds that 
    $$\cD(\bfc_n,t) = \bigcup_{\sigma\in\Sigma_{\bfc_n,t}} \sigma \circ \cD((\bfc_n)_{j_{\sigma}+1:n},t-j_{\sigma}+1).$$
    Therefore,
    \begin{align*}
    \cD(\bfc_n,t) &= \left(\bigcup_{i=1}^{m}\bigcup_{\sigma \in c_{i}} \sigma \circ \cD((\bfc_n)_{(i+1):n},t-i+1)\right) \\
    &\cup \left(\bigcup_{\sigma \in (c_{m+1}\setminus c_{1})} \sigma \circ \cD((\bfc_n)_{(m+2):n},t-m)\right).
    \end{align*}
    Note that if $w \mid q$, then $c_{m+1} \setminus c_{1} = \emptyset$, and thus the second expression is omitted.

    Next, using the ordering on $\Sigma_q$, and since the result of a deletion at the beginning of a periodic sequence is still a periodic sequence,
    \begin{align*}
    \cD(\bfc_n,t) &= \left(\bigcup_{i=1}^{m}\bigcup_{j=w(i-1)}^{wi-1} \sigma_j \circ \cD(\bfc_{n-i},t-i+1)\right) \\
    &\cup \left(\bigcup_{\substack{\{wm \leq j < w(m+1)\\\colon j<q\}}} \sigma_j \circ \cD(\bfc_{n-m-1},t-m)\right),
    \end{align*}
    where again if $w \mid q$, the second expression is ignored.
    Next, all the first $m$ symbols in $\bfc_n$ are distinct and of size $w$. 
    Moreover, the $m+1$-th symbol contains the last $q-w\cdot m$ characters of $\Sigma_q$ (could be zero if $w \mid q$), that did not appear in the first $m$ symbols of $\bfc_n$.
    Thus, we conclude that,
    \begin{align*}
        |\cD(\bfc_n,t)| &= \sum_{i=1}^{m}\sum_{j=w(i-1)}^{wi-1} |\cD(\bfc_{n-i},t-i+1)| \\
        &+ \sum_{\substack{\{wm \leq j < w(m+1)\\\colon j<q\}}} |\cD(\bfc_{n-m-1},t-m)|.
    \end{align*}
    And then,
    \begin{align*}
        |\cD(\bfc_n,t)| &= w \cdot \sum_{i=1}^{m}|\cD(\bfc_{n-i},t-i+1)| \\ 
        &+ (q-w\cdot m) \cdot |\cD(\bfc_{n-m-1},t-m)|.
    \end{align*}
    which completes the proof.
\end{IEEEproof}

Next, we show that the periodic complex sequence has the largest sub-instance ball size.

\begin{theorem}\label{theorem:maximal-sub-instance-ball}
    Let $1 \leq t \leq n-1$ and $\bfs\in\Psi_{q,w}^{n}$. Then,
    $$|\cD(\bfs,t)| \leq |\cD(\bfc_n,t)|\,.$$
\end{theorem}
\begin{IEEEproof}
    The proof is by induction on $n+t$, where the base for $n=1$ and $t=0$ is trivial, since every length-$1$ sequence is periodic, hence an equality.
    
    Let $\bfs=(\bfpsi_1,\bfpsi_2,\ldots,\bfpsi_n)$ and assume without loss of generality that $\Sigma_q$ is ordered according to the appearance in $\bfs$ (and in case of ties, choose randomly), and denote $\bfc_n=(c_1,c_2,\ldots,c_n)$ accordingly.
    Using the notations from \Cref{lemma:partitioning-the-sub-instance-ball}, where the index $j^{\bfs}_i$ denotes the first symbol in $\bfs$ such that $\sigma_i\in\bfpsi_{j^{\bfs}_i}$, while $j^{\bfs}_{i}=n$ if there is no proper index in which $\sigma$ appears in $\bfs$.
    
    Finally, let $j^{\bfc_n}_i$ as the smallest index $k$, such that $\sigma_i \in c_k$, and hence $j^{\bfs}_{\sigma} \geq j^{\bfc_n}_i$.
    Next, using \Cref{lemma:partitioning-the-sub-instance-ball},
    $$\cD(\bfs,t) = \bigcup_{i=1}^{q} \sigma_{i}\circ\cD(\bfs_{(j^{\bfs}_i+1):n},t-j^{\bfs}_i+1)\,.$$
    Consequently, and using the induction hypothesis,
    \begin{align*}
        |\cD(\bfs,t)| &= \sum_{i=1}^{q}|\cD(\bfs_{(j^{\bfs}_i+1):n},t-j^{\bfs}_i+1)| \\
        &\leq \sum_{i=1}^{q}|\cD(\bfc_{n-j^{\bfs}_i},t-j^{\bfs}_i+1)|\,.
    \end{align*}
     Since $j^{\bfs}_{\sigma} \geq j^{\bfc_n}_i$, it holds that
    \begin{align*}
        \sum_{i=1}^{q}|\cD(\bfc_{n-j^{\bfs}_i},t-j^{\bfs}_i+1)| 
        \leq \sum_{i=1}^{q}|\cD(\bfc_{n-j^{\bfc_n}_i},t-j^{\bfc_n}_i+1)| \,.
    \end{align*}
    Finally, we know that $\bfc$ has exactly $w$ distinct elements of $\Sigma$ in each of its first $m$ entries, hence the same value of $j^{\bfc_n}_i$, and that $\bfc_{m+1}$ has the remaining $q-w\cdot m$.
    Therefore, using \Cref{lemma:recursive-formula-of-t-sub-instance-ball-for-cyclic}, the proof is complete.
\end{IEEEproof}

To compute $|\cD(\bfc_n,t)|$, one can use a mapping, which is a generalization of~\cite{Hirschberg1999}.
Before presenting the formal definitions, we illustrate the main concept through a simple example.

\begin{example}\label{example:the-skipping-vector-notation}
    Let $q=4$ and $w=2$. Consider the vector
    $$\bfc_8=(\{0,1\},\{2,3\},\ldots,\{0,1\},\{2,3\}).$$
    Take two example sequences, $\bfx=(0,2,1,3)$ and $\bfy=(2,3,2,3)$. We demonstrate how each of these sequences can be represented by a vector indicating the number of ``skips'' required between symbols to extract them from $\bfc_8$ as sub-instances, provided these ``skips'' are smaller than $q$.
    First, sequence $\bfx$ can be obtained by selecting symbols as follows:
    $$(\{\underline{0},1\},\{\underline{2},3\},\{0,\underline{1}\},\{2,\underline{3}\},\ldots).$$
    Specifically, choose the first symbol (no skip) from each of the first two sets, and the second symbol (one skip) from the next two sets. Thus, $\bfx$ is mapped to the vector
    $$\waveopright{\bfx} = (0,0,1,1).$$
    Similarly, sequence $\bfy$ is obtained by selecting symbols as:
    $$(\{0,1\},\{\underline{2},3\},\{0,1\},\{2,\underline{3}\},\{0,1\},\{\underline{2},3\},\{0,1\},\{2,\underline{3}\}),$$
    which maps to
    $$\waveopright{\bfy} = (2,3,2,3).$$
    Lastly, observe that this mapping is invertible.
\end{example}

For the formal definition, let $\{\sigma_0,\sigma_1,\ldots,\sigma_{q-1}\}$ be an ordering of $\Sigma_q$. Moreover, let $\bfc_n=(c_1,c_2,\ldots,c_n)$ be a periodic complex sequence of length $n$ over $\Psi_{q,w}$. 
Then, we say that the \emph{first element} of $c_{j}=\{\sigma_{(w(j-1))\Mod{q}},\allowbreak \ldots,\sigma_{(wj-1)\Mod{q}}\}$ is $\sigma_{(w(j-1))\Mod{q}}$, although it is an element of an unordered set, since using the ordering of $\Sigma_q$ will be important next.

\begin{definition}\label{def:skipping-vectors-set}
    Let $V$ be a subset of $(n-t)$-length integer vectors over $\{ 0, 1, \ldots, q-\!1 \}$, such that for each $(v_1, v_2, \ldots\!, v_{n-t})\in V$,
    $$\sum_{j=1}^{n-t}\left\lfloor \frac{v_j}{w} \right\rfloor \leq t.$$
    We call the set $V$ the set of \emph{skipping-vectors}.
\end{definition}

\begin{definition}\label{def:mapping-of-seq-to-int-vec}
    Let $\bfT \colon \cD(\bfc_n,t) \to V$ be defined as follows.
    For $\bfx=(x_1,x_2,\ldots,x_{n-t}) \in \cD_{q,w}(\bfc_n,t) \subseteq \Sigma_q^{n-t}$, define
    $\bfv = (v_1, v_2, \ldots\!, v_{n-t})\in V$ recursively in the following steps.
    First, let $v_1$ be the number of indices skipped between $\sigma_0$ and $x_1$ with respect to the ordering of $\Sigma$, i.e., if $x_1=\sigma_j$, then $v_1=j$ (note that $j$ is always nonnegative).
    
    Also, let $k_1=\left\lfloor v_1/w \right\rfloor$, i.e., the index in $\bfc_n$ such that $x_1\in c_{k_1}$.
    Next, for all $1 < \ell \leq n-t$, let $\sigma$ be the first element in $c_{k_{\ell-1}+1}$, where $c_{k_{\ell-1}}$ is the previously matched complex symbol, and $c_{k_{\ell-1}+1}$ is therefore the following complex symbol in $\bfc_n$.
    Then, let $v_{\ell}$ be defined to be the number of indices skipped $\Mod{q}$ between $\sigma_{\ell}$ and $x_{\ell}$ with respect to the ordering of $\Sigma$.
    We will call the mapping $\bfT$ the \emph{skipping mapping}, and denote $\bfT(\bfx)\coloneqq\waveopright{\bfx}$.
\end{definition}

Note that if $\sum_{j=1}^{n-t}\left\lfloor v_j/w \right\rfloor < t$, then $\bfv \in V$ represents a sequence $\bfx \in \cD_{q,w}(\bfc_n,t)$ in which some of the last symbols of $\bfc_n$ are deleted.
\Cref{example:mapping-of-seq-to-int-vec} is provided for better understanding of \Cref{def:skipping-vectors-set} and \Cref{def:mapping-of-seq-to-int-vec}.
Furthermore, utilizing \Cref{def:mapping-of-seq-to-int-vec} one can address the problem of computing $\left|\cD_{q,w}(\bfc_n,t)\right|$, as explained in \Cref{lemma:skipping-mapping-is-invertible}.

\begin{example}\label{example:mapping-of-seq-to-int-vec}
    Let $\Sigma_7=\{0,1,\ldots,6\}$ and $\bfc_{9}\in\Psi_{7,2}$ be
    \begin{dmath*}
    \bfc_{9} = (\{0,1\},\allowbreak\{2,3\},\allowbreak\{4,5\},\allowbreak\{6,0\},\allowbreak\{1,2\},\allowbreak\{3,4\},\allowbreak\{5,6\},\allowbreak\{0,1\},\allowbreak\{2,3\}).
    \end{dmath*}
    Then, let $\bfx=(2,0,2,3) \in \cD_{7,2}(\bfc_{9},5)$ and thus $\bfT(\bfx) = (2,3,1,0)$, when
    $$
    \sum_{j=1}^{4}\left\lfloor \frac{v_j}{2} \right\rfloor = 1+1+0+0=2 < 5 = t.
    $$
    This implies that $5-2=3$ symbols were deleted at the end of $\bfc_9$, which is indeed correct.
    
    Moreover, let $\bfv=(3,6,3,1)\in V$, where 
    $$
    \sum_{j=1}^{4}\left\lfloor \frac{v_j}{2} \right\rfloor =1+3+1=5=t.
    $$
    Then, tracing back will yield 
    $$\bfy=(3,3,1,3) \in \cD_{7,2}(\bfc_{9},5),$$
    where, as expected, no deletion occurs at the end of $\bfc_{9}$.
\end{example}

\begin{lemma}\label{lemma:skipping-mapping-is-invertible}
The skipping mapping $\bfT$ is invertible.
\end{lemma}
\begin{IEEEproof}
    The proof follows from the fact that every $\bfv \in V$ can be traced back to a specific $\bfx$ by \Cref{def:mapping-of-seq-to-int-vec}.
    Let $\bfv \in V$, and $\bfc_n$ be a periodic complex sequence.
    Next, we construct a unique $\bfx$ such that $\bfT(\bfx)=\bfv$.
    For $v_1=j$, $x_1=\sigma_j$.
    Then, $x_\ell$ is determined by inverting the process described in \Cref{def:mapping-of-seq-to-int-vec}, which is unique since $v_j$ is at most $q-1$, and therefore $x_\ell$ can be of the $q$ elements in $\Sigma_q$.
    Moreover, since $\sum_{j=1}^{n-t}\left\lfloor \frac{v_j}{w} \right\rfloor \leq t$, the process can ``skip'' at most $t$ symbols of $\bfc_n$ during the choosing of the element to assign to $x_{\ell}$.
    To conclude, for every $\bfv \in V$ there exists a unique $\bfx$ such that $\bfT(\bfx)=\bfv$.
\end{IEEEproof}

Thus, by \Cref{lemma:skipping-mapping-is-invertible}, computing $\left|\cD_{q,w}(\bfc_n,t)\right|$ can be reduced to counting the number of vectors in $V$. Next, we provide examples for \Cref{def:mapping-of-seq-to-int-vec}.
\begin{example}
Let $\Sigma_6, w=4$ and $\bfc_4$, as in \Cref{example:periodic-complex-sequence}. 
Let $t=1$, then $\bfT((1,0,0))=(1,2,4)$, and $\sum_{j=1}^{3}\left\lfloor v_j/w \right\rfloor = t$, since the deletion was not at the end.
Moreover, $\bfT^{-1}((3,3,3)) = (3,1,5)$ and $\sum_{j=1}^{3}\left\lfloor v_j/w \right\rfloor < t$, as expected for a deletion of the last symbol.
\end{example}

Another recursive expression, in addition to \Cref{lemma:recursive-formula-of-t-sub-instance-ball-for-cyclic}, is presented next.
\begin{theorem}
Let $\cD_{q,w}(\bfc_n,t)$ denote the $t$-sub-instance ball for $\bfc_n \in \Psi_{q,w}^{n}$,
\begin{align*}
    \left|\cD_{q,w}(\bfc_n,t)\right| \!=\! \sum_{i=0}^{t} \left( w^{n-t-i} \cdot \binom{n\!-\!t}{i} \cdot |\cD_{q-w,w}(\bfc_t,t\!-\!i)| \right)\!,
\end{align*}
where the base case is $\cD_{q,w}(\bfc_n,t)=q^{n-t}$ for all $q \leq w$.
\end{theorem}
\begin{IEEEproof}
    As we showed in \Cref{lemma:skipping-mapping-is-invertible}, the skipping mapping $\bfT$ maps $\cD_{q,w}(\bfc_n,t)$ into $V$, the skipping-vectors set, bijectively.
    Therefore, split $V$ into disjoint subsets by counting the number of entries where $v_j \geq w$.
    Next, take each subset with $i$ entries where $v_j \geq w$, $0 \leq i \leq t$, and extract these $i$ entries by deleting all the entries $v_{k} < w$, such that the result is an $i$-length vector.
    Then, decrease the value of all the entries by $w$.
    The resultant vectors are all the $i$-length vectors $(\Tilde{v}_1, \Tilde{v}_2, \ldots, \Tilde{v}_i)$ over $\{0,1,\ldots,q-w-1\}$, and $\sum_{j=1}^{i}\left\lfloor \Tilde{v}_j/w \right\rfloor \leq t-i$.
    Thus, by \Cref{lemma:skipping-mapping-is-invertible}, the number of distinct resultant $i$-length vectors is exactly $|\cD_{q-w,w}(\bfc_t,t-i)|$.
    
    For every $0 \leq i \leq t$ there are $w^{n-t-i} \cdot \binom{n-t}{i}$ copies of the same $i$-length vectors after the deletion of all $v_{k} < w$, by choosing the $i$ entries that are $\geq w$, and then assigning all the possible value combinations between $0$ and $w-1$ to the other $n-t-i$ entries, which concludes the proof.
\end{IEEEproof}

In the following theorem, we provide an enumeration of $\left|\cD_{q,w}(\bfc_n,t)\right|$, using a generating function argument that generalizes the result of~\cite{Hirschberg1999}.

\begin{theorem}\label{theorem:sub-instance-ball-by-generating-functions}
    Let $q=m \cdot w+r$ for some $0 \leq r < w$, and denote $\cD(\bfc_n,t)$ as the $t$-sub-instance ball of $\bfc_n \in \Psi_{q,w}^{n}$.
    Then, the following holds,
    \begin{align*}
        \left|\cD(\bfc_n,t)\right| = [z^n] \left(\left(\sum_{j=1}^{m}wz^{j} + rz^{m+1} \right)^{n-t} \cdot\frac{1}{1-z}\right),
    \end{align*}
    where $[z^n]$ denotes the coefficient of $z^n$ in the polynomial of $z$ that follows it.
\end{theorem}
\begin{IEEEproof}
By \Cref{lemma:skipping-mapping-is-invertible}, one can enumerate the number of vectors in $V$.
First, divide $V$ into sets $V_{\rho}$, $0 \leq \rho \leq t$, where for all $(v_1\!, v_2, \ldots\!, v_{n-t})\!\in\!V_\rho$, it holds that
$\sum_{j=1}^{n-t}\left\lfloor \frac{v_j}{w} \right\rfloor = \rho$.
Therefore, we claim that $\left(\sum_{j=1}^{m}wz^{j} + rz^{m+1} \right)^{n-t}$ is the generating function for which the coefficient of $z^{n-\rho}$ is the number of vectors in $V_\rho$.
That is, since $v_j$ can have values between $0$ and $q-1$, but only $m+1$ distinct ones, in terms of the summation of $\lfloor v_j/w \rfloor$. Moreover, there are $m$ distinct sets of values of $v_j$ that are of size $w$, and the last set of values is of size $r$.
Note that, in terms of $\bfc_n$, one can interpret the generating function by viewing $\rho$ as the total number of skips, where a skip of size $k$ occurs when $\lfloor v_j/w \rfloor = k$.

Next, since $|V| = \sum_{\rho=0}^{t} |V_{\rho}|$, it follows that,
\begin{align*}
    |V| &= \sum_{\rho=0}^{t} [z^{n-\rho}] \left(\sum_{j=1}^{m}wz^{j} + rz^{m+1} \right)^{n-t} \\
    &= \sum_{\rho=0}^{t} [z^n] \left(\sum_{j=1}^{m}wz^{j} + rz^{m+1} \right)^{n-t}z^{\rho} \\
    &= [z^n] \left(\sum_{j=1}^{m}wz^{j} + rz^{m+1} \right)^{n-t}\sum_{\rho=0}^{t}z^{\rho},
\end{align*}
and since $[z^n] \left(\sum_{j=1}^{m}wz^{j} + rz^{m+1} \right)^{n-t}\sum_{\rho=t+1}^{\infty}z^{\rho} = 0$,
\begin{align*}
    |V| &= [z^n] \left(\sum_{j=1}^{m}wz^{j} + rz^{m+1} \right)^{n-t}\sum_{\rho=0}^{\infty}z^{\rho} \\
    &= [z^n] \left(\sum_{j=1}^{m}wz^{j} + rz^{m+1} \right)^{n-t} \frac{1}{1-z},
\end{align*}
which completes the proof.
\end{IEEEproof}

Finally, we prove the following theorem, which emphasizes a strong connection between the traditional deletion ball and the sub-instance ball if $q$ is a multiple of $w$.

\begin{theorem}\label{theorem:sub-instance-ball-of-periodic}
    Let $n>0$ and $0 \leq t < n$, and let $\bfc_n\in\Psi_{q,w}^{n}$ be the periodic complex sequence, where $w \mid q$. 
    Set $m=\frac{q}{w}$ and let $\bfp_n \coloneqq (\bfp_m)_{1:n}\in\Sigma_{m}^{n}$ be the (non-complex) periodic sequence of length $n$ over $\Sigma_{m}$.
    Then,
    $$|\cD(\bfc_n,t)| = |D(\bfp_n,t)| \,\cdot\, w^{n-t}\,.$$
\end{theorem}
\begin{IEEEproof}
    We provide a proof by induction on $n+t$.
    First, if $n+t=1$, then $n=1, t=0$ and $\cD(\bfc_n,t)$ is one complex symbol of size $w$, and $|D(\bfp_n,t)|=1$, as requested.
    
    Next, for the induction step, we start by using \Cref{lemma:recursive-formula-of-t-sub-instance-ball-for-cyclic},
    \begin{align*}
        |\cD(\bfc_n,t)| = w \cdot \sum_{i=1}^{m}|\cD(\bfc_{n-i},t-i+1)|\,.
    \end{align*}
    Then, by the inductive hypothesis
    \begin{align*}
        |\cD(\bfc_n,t)| &= w \cdot \sum_{i=1}^{m}|D(\bfp_{n-i},t-i+1)| \cdot w^{n-t-1} \\
        &= w^{n-t} \cdot \sum_{i=1}^{m}|D(\bfp_{n-i},t-i+1)| \,.
    \end{align*}
    And finally, using~\cite{Hirschberg1999},
    $|\cD(\bfc_n,t)| = w^{n-t} \cdot |D(\bfp_n,t)|,$
    which completes the proof.

    Alternatively, this follows directly from \Cref{theorem:sub-instance-ball-by-generating-functions}.
    Here, $r=0$, thus the expression is reduced into
    \begin{align*}
        \left|\cD_{q,w}(\bfc_n,t)\right| = w^{n-t} \cdot [z^n] \left(\sum_{j=1}^{m}z^{j} \right)^{n-t} \frac{1}{1-z},
    \end{align*}
    where~\cite{Hirschberg1999} proved that
    $$|D(\bfp_n,t)| = [z^n] \left(\sum_{j=1}^{m}z^{j} \right)^{n-t} \frac{1}{1-z},$$
    which completes the proof.
\end{IEEEproof}

\subsection{Information Rate Analysis}\label{subsec:information-rate-analysis}

We aim to completely solve \Cref{problem:optimal-information-rate}, and compute the values $f(q,w)$ and $f(q,w,\alpha)$.
Using \Cref{theorem:maximal-sub-instance-ball} and \Cref{lemma:number-of-sub-instances},  it follows that $f(q,w,\alpha) = \cR_{q,w}(\bfc_n,\alpha)$ and similarly $f(q,w) = \cR_{q,w}(\bfc_n)$.
Based on the results of \Cref{subsec:sub-instance-ball}, we provide explicit expressions for $f(q,w)$ and $f(q,w,\alpha)$, and relate them to the results of~\cite{Lenz2021} on $f(q,1)$ and $f(q,1,\alpha)$.

\begin{theorem}\label{theorem:max-star-information-rate}
Let $q,w$ be positive integers such that $w \leq q$.
\begin{align*}
    f(q,w) = \max_{0 < \alpha < 1} f(q,w,\alpha).
\end{align*}
\end{theorem}
\begin{IEEEproof}
    Since $f(q,w) = \cR_{q,w}(\bfc_n)$,
    \begin{align*}
        \cR_{q,w}(\bfc_n) = \limsup_{t\to\infty} \frac{\log_2\left( \sum_{i=0}^{t}|\cD(\bfc_n,i)|\right)} {t} \,.
    \end{align*}
    Let $i^{*}$ be the index for which $|\cD(\bfc_n,i^{*})|$ is the largest. 
    From \Cref{lemma:number-of-sub-instances}, the maximum over $\alpha$
    is achieved when $\alpha^{*}$ satisfies $i^{*} = t-\lfloor\alpha^{*} t\rfloor$. 
    Hence,
    \begin{align*}
        &\limsup_{t\to\infty} \frac{\log_2\left( \sum_{i=0}^{t}|\cD(\bfc_n,i)|\right)} {t} \\
        & =\limsup_{t\to\infty} \frac{\log_2(|\cD(\bfc_n,i^{*})|) + \log_2\left( \sum_{i=0}^{t}\frac{|\cD(\bfc_n,i)|}{|\cD(\bfc_n,i^{*})|}\right)} {t} \\
        & =\limsup_{t\to\infty} \frac{\log_2(|\cD(\bfc_n,i^{*})|)} {t}
        + \frac{O(\log(t))} {t} \\
        & =\limsup_{t\to\infty} \frac{\log_2(|\cD(\bfc_n, t-\lfloor\alpha^{*} t\rfloor)|)} {t} \\
        &=\max_\alpha \cR_{q,w}(\bfc_n,\alpha),
    \end{align*}
    which completes the proof.
\end{IEEEproof}

A result for computing the value of $f(q,w)$ is given next.

\begin{theorem}\label{theorem:f(q-w)-using-singularity}
    For any $q$ and $w$ such that $q=mw+r$ for some $0 \leq r < w$, we have
    $$f(q,w) = -\log_2(z_{q,w}),$$
    where $z_{q,w}$ is the unique positive solution of 
    $$\sum_{j=1}^{m}wz^j+rz^{m+1}=1.$$
\end{theorem}
\begin{IEEEproof}
    We know from \Cref{theorem:sub-instance-ball-by-generating-functions} that
    \begin{align*}
        \left| \cD(\bfc_n,t) \right| = [z^n] \left(\sum_{j=1}^{m}wz^{j} + rz^{m+1} \right)^{n-t} \cdot\frac{1}{1-z}.
    \end{align*}
    Thus,
    \begin{align*}
         N(\bfc_n, n) &= \sum_{t=0}^{n} |\cD(\bfc_n,t)| \\
         &= \sum_{t=0}^{n} [z^n] \frac{1}{1-z} \left(\sum_{j=1}^{m}wz^{j} + rz^{m+1} \right)^{n-t} \\
         &= [z^n] \frac{1}{1-z} \sum_{t=0}^{n} \left(\sum_{j=1}^{m}wz^{j} + rz^{m+1} \right)^{t} \\
         &= [z^n] \frac{1}{1-z} \sum_{t=0}^{\infty} \left(\sum_{j=1}^{m}wz^{j} + rz^{m+1} \right)^{t} \\
         &= [z^n] \frac{1}{1-z} \left( 1 - \sum_{j=1}^{m}wz^{j} + rz^{m+1} \right)^{-1}.
    \end{align*}
    Applying~\cite[Theorem IV.7]{Flajolet2009}, the coefficient $[z^n]$ is asymptotically $R^{-n}$, where $R$ is the smallest singularity of 
    $$\frac{1}{1-z} \left( 1 - \sum_{j=1}^{m}wz^{j} + rz^{m+1} \right)^{-1},$$
    i.e., $R$ is the unique solution to $\sum_{j=1}^{m}wz^j+rz^{m+1}=1$, which is denoted as $z_{q,w}$.
    Thus, $\lim_{n\to\infty} N(\bfc_n, n) = (z_{q,w})^{-n}$ and by \Cref{lemma:number-of-sub-instances} and \Cref{theorem:maximal-sub-instance-ball}, it holds that
    $$
        f(q,w) = \limsup_{n\to\infty}\frac{\log_2\left(N(\bfc_n, n)\right)}{n}
        = -\log_2(z_{q,w}).
    $$
\end{IEEEproof}

Note that \Cref{theorem:f(q-w)-using-singularity} coincides with~\cite[Theorem 3]{Lenz2020} when $w=1$. 
The following is a generalization of \Cref{theorem:f-for-w=1}.
\begin{theorem}\label{theorem:max-information-rate}
Let $q$ and $w$ be integers such that $q=mw+r$ for some $0 \leq r < w$ and $0 \leq \alpha < 1$.
Then, $f(q,w,\alpha)=$
\begin{align*}
    \begin{cases}
    \alpha \log_2(q) &\, 0 \leq \alpha < b\,, \\
    \alpha\log_2\left(\sum_{i=1}^{m} w \!\cdot\! (x_{\alpha})^{i-\frac{1}{\alpha}} +r \!\cdot\! (x_{\alpha})^{m+1-\frac{1}{\alpha}} \right) &\, b < \alpha < 1,
    \end{cases}
\end{align*}
where $b \coloneqq \frac{q}{\left(m+1\right)\left(q-\frac{wm}{2}\right)},$
and $x_{\alpha}$ is the unique solution on the interval $0<x<1$ to 
$$\sum_{i=1}^{m} w\left(1-\alpha i\right)x^{i} + r\left(1-\alpha\left(m+1\right)\right)x^{m+1}=0.$$
\end{theorem}
\begin{IEEEproof}
Without loss of generality, $\Sigma_q=\{0,1,\ldots,q-1\}$, and denote $\beta=q/\gcd(q,w)$.
First, we define a constraint graph $G_{q,w}$ for the synthesis using the synthesis periodic complex sequence $\bfc_n$. 
Let $\{\psi_j\}_{j=1}^{\beta}$ be the complex symbols in $\bfc_n$, with respect to the order in which they appear in $\bfc_n$. 
Accordingly, $\{v_j\}_{j=1}^{\beta}$ are the vertices of $G_{q,w}$ as their representatives.
Each vertex has $q$ outgoing edges.
An edge from $v_i$ to $v_j$ is labeled with the character $k \in \Sigma_q$ if $k \in \psi_j$ and $k \notin \psi_{h}$ for all $i < h < j \Mod{\beta}$, and its cost is $j-i$. i.e., there is a $k$-labeled edge from $v_i$ to $v_j$ if $\psi_j$ is the following complex symbol (after $\psi_i$, not including it) in $\bfc_n$ that has the element $k$.
Note that for each vertex, there are exactly $w$ edges with the costs $1,2,\ldots,m$ ($w$ edges for each), and $r$ edges with the cost $m+1$.
As in~\cite{Lenz2021}, the graph $G_{q,w}$ is strongly connected, deterministic, and cost-diverse.
Using~\cite[Theorem III.1]{Lenz2021} and identifying 
$$\rho_{G_{q,w}}(x)=\sum_{i=1}^{m}wx^{i}+rx^{m+1},$$
by applying~\cite[Lemma III.11]{Lenz2021}, we deduce the result.

\end{IEEEproof}

Next, we show that when $q$ is a multiple of $w$, it provides a direct connection to the results of Lenz et al.~\cite{Lenz2020, Lenz2021}.

\begin{theorem}\label{theorem:max-information-rate-w-divides-q}
Let $q,w$ be positive integers such that $w \mid q$, and $m\coloneqq \frac{q}{w}$. Then, for all $0 \leq \alpha \leq 1$, it holds that
$$f(q,w,\alpha) = \alpha \log_2(w) + f(m,1,\alpha).$$
\end{theorem}
\begin{IEEEproof}
    Using \Cref{lemma:number-of-sub-instances}, \Cref{theorem:maximal-sub-instance-ball} and \Cref{theorem:sub-instance-ball-of-periodic},
    \begin{align*}
        f(q,w,\alpha) &= \limsup_{t\to\infty} \frac{\log_2\left( |\cD(\bfc_t,t-\lfloor \alpha t \rfloor)| \right)} {t} \\
        &= \limsup_{t\to\infty} \frac{\log_2\left( |\bfD(\bfp_t,t-\lfloor \alpha t \rfloor)| \cdot w^{\lfloor \alpha t \rfloor}\right)} {t},
    \end{align*}
    where $\bfp_t \in \Sigma_m^{t}$ is the periodic (alternating) sequence. Then,
    \begin{align*}
        &= \alpha \log_2(w) + \limsup_{t\to\infty} \frac{\log_2\left(|\bfD(\bfp_t,t-\lfloor \alpha t \rfloor)|\right)} {t}.
    \end{align*}
    Applying~\cite{Lenz2020},
    \begin{align*}
        = \alpha \log_2(w) + \max_{\bfx\in\Sigma_m^{*}} \cR(\bfx,\alpha)= \alpha \log_2(w) + f(m,1,\alpha).
    \end{align*}
\end{IEEEproof}

Note that $f(m,1,\alpha)=\max_{\bfx\in\Sigma_{m}^{*}}\cR(\bfx,\alpha)$ is the information rate in the classical case, as defined in~\cite{Lenz2020}.
Thus, \Cref{problem:optimal-information-rate} is solved.
Moreover, using \Cref{theorem:f-for-w=1}, one can derive the explicit expression, as summarized in ~\Cref{corollary:max-cap-in-alpha-explicit}.
\begin{corollary}\label{corollary:max-cap-in-alpha-explicit}
Let $q,w>0$ be integers such that $w \mid q$, and $m\coloneqq\frac{q}{w}$. It holds that
\begin{align*}
    f(q,w,\alpha)\!=
    \begin{cases}
       \alpha \log_2\left( w\alpha\sum_{i=1}^{m}i\!\cdot\!(x_{m,\alpha})^{i-\frac{1}{\alpha}} \right) &, \frac{2}{m+1} \!\leq\! \alpha, \\
        \alpha \log_2(q)&,  \alpha \!<\! \frac{2}{m+1}.
    \end{cases}
\end{align*}
where $x_{m,\alpha}$ is the unique solution to $\sum_{i=1}^{m} (1-\alpha i)x^{i}=0$ on the interval $0<x<1$.
\end{corollary}
Finally, for the practical use of the DNA alphabet, consider the following corollary.
\begin{corollary}\label{corollary:q=4-w=2-cap}
Let $q=4$ and $w=2$. It holds that
\begin{align*}
    f(4,2,\alpha) &= 
    \begin{cases}
        \alpha + \alpha h( \alpha^{-1} - 1 ), & \frac{2}{3} \leq \alpha\,,\\
        2\alpha, & \text{otherwise}
    \end{cases}
    \,,
\end{align*}
and that the maximum is attained at $\alpha \approx 0.789$. Then,
$f(4,2) \approx f(4,2,0.789) = 1.45.$
\begin{IEEEproof}
    Follows directly from \Cref{theorem:f-for-w=1}, \Cref{theorem:max-information-rate}, and \Cref{theorem:max-star-information-rate}.
\end{IEEEproof}
\end{corollary}

In~\Cref{table:summary-f-4-w}, we summarize the results for $f(4,w)$ based on Lenz~\cite{Lenz2020}, \Cref{corollary:q=4-w=2-cap} and the usage of \Cref{theorem:f(q-w)-using-singularity}.
A comprehensive summary of the results is presented in~\Cref{table:summary-f-q-w-info-rate}.

\begin{table}[H]
\centering
\caption{Values of $f(4,w)$, rounded to 3 decimal places.}
\begin{tabular}{|| c || c | c | c | c ||}
\hline
$w$ & $1$ & $2$ & $3$ & $4$ \\ [0.5ex] \hline\hline
$f(4,w)$ & $0.947$ & $1.450$ & $1.724$ & $2.000$ \\ \hline
\end{tabular}
\label{table:summary-f-4-w}
\end{table}


Finally, we wish to analyze the asymptotic behavior of $f(q,w)$, and begin with the following theorem as an upper bound, which will be followed by a proof that it is asymptotically obtainable.

\begin{theorem}\label{theorem:upper-bound-on-information-rate}
    Let $q,w$ be positive integers.
    \begin{align*}
        f(q,w) \leq \log_2(w+1).
    \end{align*}
\end{theorem}
\begin{IEEEproof}
    Let $\bfs \in \Psi_{q,w}^{*}$ be a complex synthesis sequence that maximizes $f(q,w)$.
    Thus, after $\tau>0$ cycles, each strand $\bfx\in\Sigma_{q}^{\leq \tau}$ that can be synthesized by $\bfs$ can be injectively mapped into a vector $\bfx^{\prime} \in \Sigma_{w+1}^{\tau}$, where the $j$-th entry in $\bfx^{\prime}$ is a character in $\bfs_j$ (which is a complex symbol) if this character is being added to $\bfx$ in the $j$-th cycle, or the $j$-th entry in $\bfx^{\prime}$ is $\varepsilon$ otherwise. 
    Thus, at most, $\bfs\in\Psi_{q,w}^{*}$ can synthesize all the sequences in $\Sigma_{w+1}^{\tau}$ after $\tau$ cycles, and thus $N(\bfs,\tau) \leq (w+1)^\tau$ and $f(q,w) \leq \log_2(w+1)$.
\end{IEEEproof}

\begin{figure}
\centering
\includegraphics[width=\linewidth]{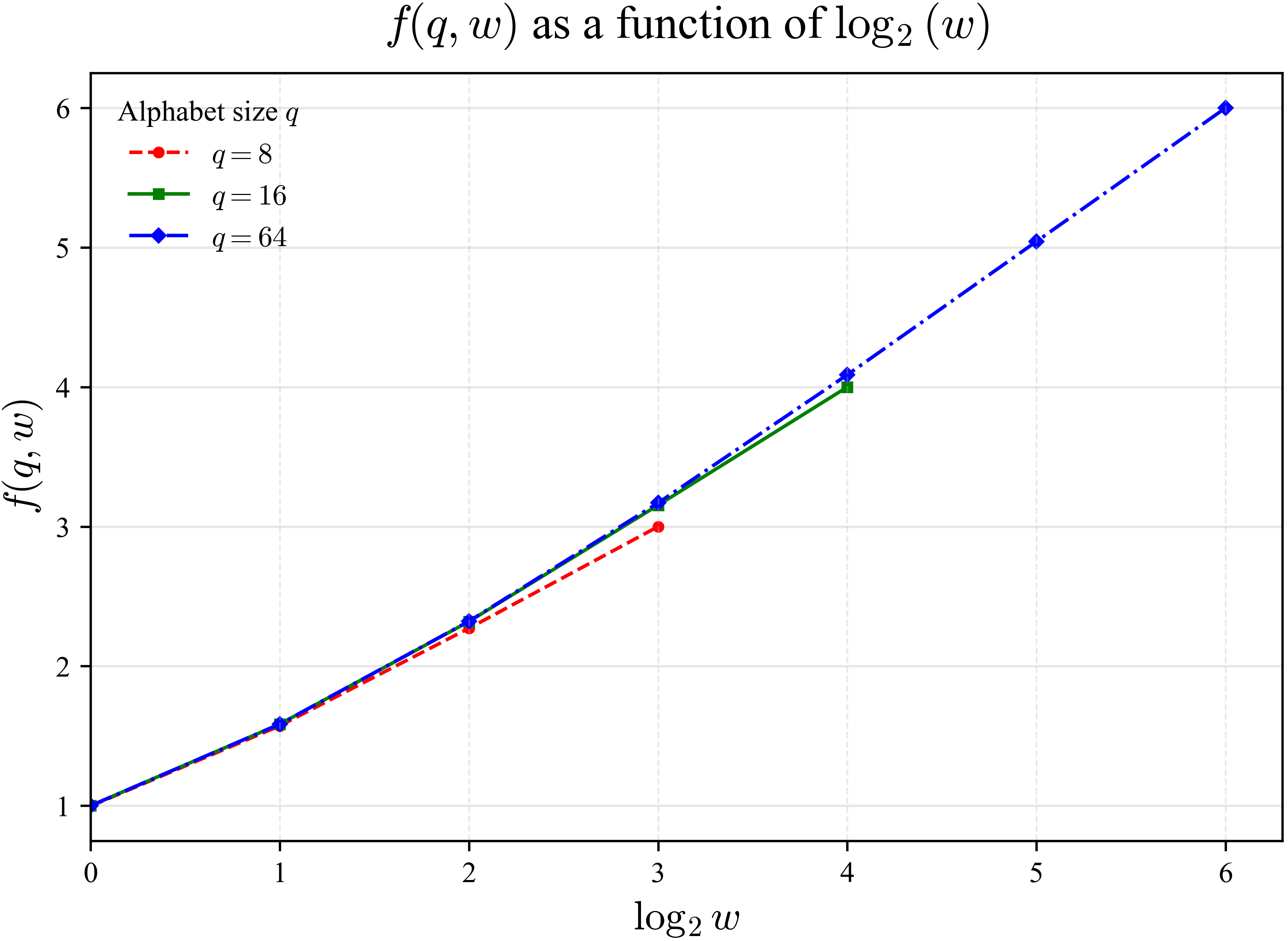}
\caption{The values of $f(q,w)$ for $q=8,16,64$ all $w\leq q$ that divides $q$. The values of $w$ are in logarithmic scale and the dashed lines between the values of $\log_2(w)$ is added for clarity.}
\label{fig:graph-of-w}
\end{figure}

In \Cref{table:summary-f-q-w-info-rate} we provide the values of $f(q,w)$ for $q$ and $w$ which are selected powers of $2$. 
We also provide the value of $\argmax_{0 < \alpha < 1} f(q,w,\alpha)$ that yields $f(q,w)$, according to \Cref{theorem:max-star-information-rate}.
In \Cref{fig:graph-of-w} one can observe the rate at which $f(q,w)$ converges to $\log_2(q)$, and that rate converges towards $\log_2(w+1)$ when $q\to\infty$, as can be seen in \Cref{table:summary-f-q-w-info-rate} as well, and that $\argmax_{0 < \alpha < 1} f(q,w,\alpha)$ converges to $\frac{w}{w+1}$ when $q\to\infty$.

Next, we capture these observations in the following formal statement.

\begin{theorem}\label{theorem:information-rate-for-infinite-q}
    Let $w$ be a positive integer. Then,
    \begin{align*}
        \lim_{q\to\infty} f\left(q,w,\frac{w}{w+1}\right) = \log_2(w+1).
    \end{align*}
    Thus, 
    \begin{align*}
        \lim_{q\to\infty} \argmax_{0 < \alpha < 1} f\left(q,w,\alpha\right) = \frac{w}{w+1}.
    \end{align*}
    Moreover,
    \begin{align*}
        \lim_{q\to\infty} f\left(q,w\right) = \log_2\left(w+1\right).
    \end{align*}
\end{theorem}
\begin{IEEEproof}
    Since $w$ is fixed and $q\to\infty$, also $m\to\infty$. Then, by choosing $\alpha=\frac{w}{w+1}$, we first need to find the root $x_{m,\alpha}$, i.e., $x_{\infty,\frac{w}{w+1}}$, to use the results of \Cref{corollary:max-cap-in-alpha-explicit}.
    Note that since $f(q,w,w/(w+1))$ is a non-decreasing function, it is sufficient to look at a subsequence of $q$ which are multiples of $w$.
    Thus, we begin with the series,
    $$\sum_{i=1}^{\infty} \left(1-\frac{w}{w+1}\cdot i \right) x^{i}.$$
    Since we look for a root on the interval $0<x<1$, this power series converges as follows
    \begin{align*}
        \sum_{i=1}^{\infty} \left(1-\frac{w}{w+1}\cdot i\right)x^{i} &= -\frac{x (x + \frac{w}{w+1} - 1)}{{(x-1)}^{2}} \\
        &= -\frac{x ((w+1)x + w - w -1)}{{(w+1)(x-1)}^{2}} \\
        &= -\frac{x ((w+1)x - 1)}{{(w+1)(x-1)}^{2}}.
    \end{align*}
    Therefore, the root on the interval $0<x<1$ is obtained as the solution of 
    $$-\frac{x ((w+1)x - 1)}{{(w+1)(x-1)}^{2}} = 0,$$
    Consequently, $x_{\infty,\frac{w}{w+1}}=\frac{1}{w+1}$.
    Finally,
    \begin{align}
        &\lim_{q\to\infty} f\left(q,w,\frac{w}{w+1}\right) \nonumber \\
        &=\frac{w}{w+1} \log_2\left(w \frac{w}{w+1} \sum_{i=1}^{\infty} i \left(\frac{1}{w+1}\right)^{i-\frac{w+1}{w}} \right) \nonumber \\
        &=\frac{w}{w+1} \log_2\left( \frac{w^2}{w+1}(w+1)^{\frac{w+1}{w}} \sum_{i=0}^{\infty} i \left( \frac{1}{w+1} \right)^{i} \right) \label{eq:mid-compute-lim-f(q,w,w/w+1)}
     \end{align}
     Since $w>0$, then $0<\frac{1}{w+1}<1$, and the power series 
     $$\sum_{i=0}^{\infty} i \cdot \left(\frac{1}{w+1}\right)^{i},$$
     converges to 
     $$\frac{\frac{1}{w+1}}{(1-\frac{1}{w+1})^2}=\frac{w+1}{w^2}.$$
     Thus, continue from \eqref{eq:mid-compute-lim-f(q,w,w/w+1)},
     \begin{align*}   
        &=\frac{w}{w+1} \log_2\left( \frac{w^2}{w+1} \cdot (w+1)^{\frac{w+1}{w}} \cdot \frac{w+1}{w^2} \right) \\
        &=\frac{w}{w+1} \log_2\left((w+1)^{\frac{w+1}{w}}\right) = \log_2(w+1).
    \end{align*}
    We conclude that 
    $\lim_{q\to\infty} f\left(q,w,\frac{w}{w+1}\right) = \log_2\left(w+1\right)$.
    And using \Cref{theorem:upper-bound-on-information-rate}, 
    $$\lim_{q\to\infty} f\left(q,w\right) = \log_2\left(w+1\right),\vspace{-3pt}$$
    where the $\alpha$ that maximizes $\lim_{q\to\infty} f(q,w,\alpha)$ is $\frac{w}{w+1}$.
    
    Although \Cref{corollary:max-cap-in-alpha-explicit}, formally assumes $w \mid q$, we can apply it because $f\left(q,w,\frac{w}{w+1}\right)$ is a monotonically increasing function in $q$ for fixed $w$.
\end{IEEEproof}

\subsection{Synthesis Code Constructions}\label{subsec:codes}
This subsection generalizes the code construction introduced in~\cite{Lenz2020}.
Without loss of generality, we assume the alphabet is given by $\Sigma_q = {0,1,\ldots,q-1}$. We focus on synthesis codes for periodic complex sequences $\bfc_{T} \in \Psi_{q,w}^{T}$.

Specifically, let $T$ be a positive integer.
Our goal is to construct codes $\cC \subseteq \Sigma_{q}^{n}$ such that each sequence $\bfx \in \cC$ can be synthesized from $\bfc_n$ within $T$ synthesis cycles, where $\bfc_n$ is the periodic complex sequence.
We denote by $T_{\bfc_n}(\bfx)$ the minimal number of synthesis cycles required to synthesize $\bfx$ from $\bfc_n$.
The following example illustrates how $T_{\bfc_n}(\bfx)$ can be derived explicitly.

\begin{example}\label{example:x-prime-skips}
    Let $\bfc_8 \in \Psi_{4,2}$ and consider $\bfy$ from \Cref{example:the-skipping-vector-notation}, which demonstrates that $\bfy$ requires exactly $8$ synthesis cycles.
    
    By a slight abuse of notation, we denote the skipping vector as $\waveopright{\bfy}=(3,4,3,4)$, where each entry of the original skipping vector is increased by one to facilitate the direct computation of the number of synthesis cycles, as formalized next in \Cref{lemma:complex-synthesis-cycles-formula}, $T_{\bfc_n}(\bfy) = \left\lceil3/2\right\rceil + \left\lceil4/2\right\rceil + \left\lceil3/2\right\rceil + \left\lceil4/2\right\rceil = 8$.
    
    Additionally, we introduce the \emph{inverse skipping vector}, denoted by $\waveopleft{\bfy}$, defined as $\waveopleft{\bfy_i}=q+1-\waveopright{\bfy_i}$.
    Here, the inverse skipping vector of $\bfy$ is $\waveopleft{\bfy}=(2,1,2,1)$, corresponding to the skipping vector of the sequence $(1,2,1,2)$, which requires only $4$ synthesis cycles.
    This subsection introduces a code construction leveraging inverse skipping vectors to ensure synthesis in fewer cycles.
\end{example}

Next, we formally justify the slight abuse of notation.
We redefine $\waveopright{\bfx}$ to be $\bfT(\bfx)+1$, where $\bfT(\cdot)$ is the mapping defined in \Cref{def:mapping-of-seq-to-int-vec}, and $+1$ indicates element-wise addition.
This adjustment is the only difference between the definition of $\waveopright{\bfx}$ in \cref{subsec:sub-instance-ball} and this subsection.
Note again that $\bfx$ can be uniquely recovered from $\waveopright{\bfx}$.

\begin{lemma}\label{lemma:complex-synthesis-cycles-formula}
It holds that:
\begin{align*}
    T_{\bfc_n}(\bfx) = \sum_{i=1}^{|\bfx|}\left\lceil\frac{\waveopright{x_i}}{w}\right\rceil
\end{align*}
\end{lemma}
\begin{IEEEproof}
    The vector $\waveopright{\bfx}$ counts the number of skips ($+1$) required to reach the next synthesized element.
    Hence, $\lceil\waveopright{x_i}/w\rceil$ holds exactly the number of complex symbol skips (plus $1$) that it needs, which is the number of synthesis cycles.
\end{IEEEproof}

Furthermore, we present an explicit single-redundancy construction for the following code:
\begin{align*}
    \cC_{\lfloor b_{q,w}/2 \rfloor} = \left\{ \bfx\in\Sigma_q^n \,\colon\, \sum_{i=1}^{n}\lceil\waveopright{x_i}/w\rceil \leq \lfloor b_{q,w}/2\rfloor \right\},
\end{align*}
where,
$$b_{q,w} = 
\begin{cases}
    n\left(m+1\right) &\text{ if } w\mid q \text{ and } q=mw, \\
    n\left(1+\left\lceil\frac{q+1}{w}\right\rceil\right) &\text{ if } w\nmid q.
\end{cases}$$
For the explicit encoder, let us first define the inverse skipping vector formally by setting $\waveopleft{x_i}\in\Sigma_q$ as $q+1-\waveopright{x_i}$.
Additionally, denote by $\bfx^{\text{c}}$ the sequence whose skipping vector is exactly $\waveopleft{\bfx}$.
The explicit construction is provided in \Cref{encoder:less-synth-cycles} and \Cref{decoder:less-synth-cycles}, where $f(\cdot)$ denotes the function that computes either $\bfx$ from its skipping vector $\waveopright{\bfx}$ or $\bfx^{\text{c}}$ from $\waveopleft{\bfx}$.


\begin{encoder}
    \caption{ENC, an encoder for $\cC_{\lfloor b_{q,w}/2 \rfloor}$.}
    \label[Encoder]{encoder:less-synth-cycles}
    \begin{algorithmic}[1]
        \renewcommand{\algorithmicrequire}{\textbf{Input:}}
        \REQUIRE $\bfy\in\Sigma_{q}^{n-1}$.
        \renewcommand{\algorithmicensure}{\textbf{Output:}}
        \ENSURE $\bfc\in\cC_{\lfloor b_{q,w}/2 \rfloor}\subseteq \Sigma_{q}^{n}$.
        
        \STATE Let $\waveopright{\bfx}=(\waveopright{y_1},\waveopright{y_2},\ldots,{\waveopright{y}}_{n-1}, 1) \coloneqq (\waveopright{x_1}, \waveopright{x_2}, \ldots, \waveopright{x_n})$.
        \STATE Let $S = \sum_{i=1}^{n}\left\lceil\frac{\waveopright{x_i}}{w}\right\rceil$.
        
        \IF {$\left(S \leq \left\lfloor\frac{b_{q,w}}{2}\right\rfloor\right)$}
            \STATE $\bfx = f(\waveopright{\bfx})$.
            \STATE $\bfc = \bfx$.
        \ELSE
            \STATE $\bfx^{\text{c}} = f(\waveopleft{\bfx})$.
            \STATE $\bfc = \bfx^{\text{c}}$.
        \ENDIF
        
        \RETURN $\bfc$
    \end{algorithmic} 
\end{encoder}

\begin{decoder}
    \caption{DEC, a decoder for $\cC_{\lfloor b_{q,w}/2 \rfloor}$.}
    \label[Decoder]{decoder:less-synth-cycles}
    \begin{algorithmic}[1]
        \renewcommand{\algorithmicrequire}{\textbf{Input:}}
        \REQUIRE $\bfc\in\cC_{\lfloor b_{q,w}/2 \rfloor}$.
        \renewcommand{\algorithmicensure}{\textbf{Output:}}
        \ENSURE $\bfy\in\Sigma_{q}^{n-1}$.
        
        \STATE Compute $\waveopright{\bfc}$.
        \IF {$(\waveopright{c_n} == 1)$}
            \STATE $\bfy = (c_1, c_2, \ldots, c_{n-1})$.
        \ELSE
            \STATE $\bfc^{\text{c}} = f(\waveopleft{\bfc})$.
            \STATE $\bfy = (c_1^{\text{c}}, c_2^{\text{c}}, \ldots, c_{n-1}^{\text{c}})$.
        \ENDIF
        
        \RETURN $\bfy$.
        
    \end{algorithmic} 
\end{decoder}

Next, we provide an explanation for the correctness of the construction.

\begin{claim}
    \Cref{encoder:less-synth-cycles} outputs $\bfc\in\cC_{\lfloor b_{q,w}/2 \rfloor}$.
\end{claim}
\begin{IEEEproof}
    We will show that 
    $$
    T_{\bfc_n}(\bfx) + T_{\bfc_n}(\bfx^{\text{c}}) \leq n\left(1+\left\lceil\frac{q+1}{w}\right\rceil\right),
    $$
    and when $w \mid q$, i.e., $q=m\cdot w$, it holds that  
    $$
    T_{\bfc_n}(\bfx) + T_{\bfc_n}(\bfx^{\text{c}}) = n\left\lceil\frac{q+1}{w}\right\rceil = n(m+1).
    $$
    Therefore, either $T_{\bfc_n}(\bfx)\leq \left\lfloor\frac{b_{q,w}}{2}\right\rfloor$ or $T_{\bfc_n}(\bfx^{\text{c}})\leq \left\lfloor\frac{b_{q,w}}{2}\right\rfloor$.
    First, in the general case, it holds that
    \begin{align*}
        T_{\bfc_n}(\bfx) + T_{\bfc_n}(\bfx^{\text{c}}) &= \sum_{i=1}^{n} \left\lceil \frac{\waveopright{x_i}}{w} \right\rceil + \sum_{i=1}^{n} \left\lceil \frac{q+1-\waveopright{x_i}}{w} \right\rceil \\
        &\leq \sum_{i=1}^{n} \left\lceil \frac{\waveopright{x_i}}{w} \right\rceil + \left\lceil \frac{q+1}{w} \right\rceil + \left\lceil \frac{-\waveopright{x_i}}{w} \right\rceil \\
        &\leq n\left(1+\left\lceil \frac{q+1}{w} \right\rceil \right).
    \end{align*}
    Moreover, when $w \mid q$, denote $\waveopright{x_i} = a \cdot w + \rho,\, 0 \leq \rho < w$, and $q=m \cdot w$.
    Then, it holds that
    $$
    \left\lceil \frac{\waveopright{x_i}}{w} \right\rceil = \begin{cases}
        a &; \rho=0, \\
        a+1 &; \rho>0.
    \end{cases}
    $$
    Also, it holds that
    $$
    \left\lceil \frac{q+1-\waveopright{x_i}}{w} \right\rceil = \left\lceil \frac{mw+1-aw+\rho}{w} \right\rceil = m-a+\left\lceil \frac{1-\rho}{w} \right\rceil.
    $$
    Therefore, if $\rho=0$,
    $$
    \left\lceil \frac{\waveopright{x_i}}{w} \right\rceil + \left\lceil \frac{q+1-\waveopright{x_i}}{w} \right\rceil = a+m-a+ \left\lceil \frac{1}{w} \right\rceil = m+1.
    $$
    Otherwise, $0 < \rho < w$, which yields $-1 < \frac{1-w}{w} < \frac{1-\rho}{w} \leq 0$, and $m-a+\left\lceil \frac{1-\rho}{w} \right\rceil = m-a$.
    Thus, 
    $$
    \left\lceil \frac{\waveopright{x_i}}{w} \right\rceil + \left\lceil \frac{q+1-\waveopright{x_i}}{w} \right\rceil = a+1+m-a = m+1.
    $$
    To conclude, when $w \mid q$, it holds that
    \begin{align*}
        T_{\bfc_n}(\bfx) + T_{\bfc_n}(\bfx^{\text{c}}) &= \sum_{i=1}^{n} \left\lceil \frac{\waveopright{x_i}}{w} \right\rceil + \sum_{i=1}^{n} \left\lceil \frac{q+1-\waveopright{x_i}}{w} \right\rceil \\
        &= \sum_{i=1}^{n} \left\lceil \frac{\waveopright{x_i}}{w} \right\rceil + \left\lceil \frac{q+1-\waveopright{x_i}}{w} \right\rceil \\
        &=n(m+1).
    \end{align*}
\end{IEEEproof}

\section{Optimal Synthesis Sequence}\label{sec:SCCS}
In this section, we address \Cref{problem:shortest-synthesis-sequence} by introducing in \Cref{algorithm:SCCS} a polynomial-time algorithm that computes the length of an SCCS of a collection of sequences using dynamic programming and reconstructs an SCCS, and in \Cref{algorithm:approx-SCCS}, an approximation algorithm for the case where the number of sequences in the collection is large.

\subsection{Dynamic Programming Approach}
Our approach is a generalization of the well-studied problem of computing the shortest common supersequence (SCS)~\cite{Itoga1981}.
Let $\bfx[j]$ be the $j$-th entry of $\bfx$, and set $\bfx[0]=\varepsilon$, i.e., the empty word. For contiguous substrings, use $\bfx[i \colon j]$.
Next, we need to introduce new notations, based on~\cite{Itoga1981}, to establish the recursion behind the dynamic programming approach, which is presented in Claim~\ref{claim:SCCS-recursion}. For $\bfpsi\in\Psi_{q,w}$, $\bfx\in\Sigma_q^{*}$, and $i \geq 1$, let
\begin{align*}
    g(\bfpsi,\bfx, i) =
    \begin{cases}
        i-1, & \text{if } \bfx[i]\in\bfpsi, \\
        i, & \text{otherwise}.
    \end{cases}
\end{align*}
And assume $g(\bfpsi,\bfx, 0) = 0$ by convention. For example, 
\begin{align*}
    &g(\{A,C\}, (A,C,G), 2) = 1, \\
    &g(\{A,C\},(A,C,G), 3) = 3.
\end{align*}
For brevity, let $g(\bfpsi, \bfx_{1\to k},i_{1\to k})$ represent the following sequence, $(g(\bfpsi, \bfx_1, i_1), g(\bfpsi, \bfx_2, i_2), \ldots, g(\bfpsi, \bfx_k, i_k))$. 
Lastly, denote 
$$L[i_1,\ldots,i_k] \coloneqq |SCCS(\bfx_1[1\!:\!i_1], \bfx_2[1\!:\!i_2], \ldots, \bfx_k[1\!:\!i_k])|,$$
where $\bfx_j[1:0]$ is the empty string.

\begin{claim}\label[Claim]{claim:SCCS-recursion}
Let $\cX  = \{ \bfx_1, \ldots, \bfx_k \} \subseteq \Sigma_{q}^{*}$ be sequences of lengths $n_1,\dots,n_k$, respectively. For all $0 \leq i_j \leq n_j$, let $i_j^{\prime} = \max\{0,i_j-1\}$ and $U = \{\bfx_j[i_j] \colon i_j>0\}_{j=1}^{k}$. 
Also, let $V = \{\bfpsi\in\Psi_{q,w} \colon U \cap \bfpsi \neq \emptyset\}$ represent the valid symbols. 
Then, $L[i_1,\ldots,i_k] = $
\begin{align*}
        \begin{cases}
            0 &\text{ if } i_1=\dots=i_k=0 \\
            L[i_1^{\prime},\ldots,i_k^{\prime}] \!+\! 1 &\text{ if} \,\left| U \right| \!\leq\! w \\
            \min_{\bfpsi \in V} \{ L[g(\bfpsi,\! \bfx_{1\to k},\!i_{1\to k})]\} \!+\! 1 &\text{ otherwise,}
        \end{cases}
\end{align*}
where the minimum in the last case is taken over $\bfpsi \in \Psi_{q,w}$ such that there exists $\bfx_j\in\cX$ for which  $\bfx_j[i_j] \in \bfpsi$.
\end{claim}

\begin{remark}\label{remark:recursion-step-shortening-explained}
    Note that we only consider $\bfpsi \in V$ because, without it, $g(\bfpsi, \bfx_{1\to k},i_{1\to k})$ would return $(i_1, \ldots, i_k)$, creating a circular argument and making this case undefined.
\end{remark}

One way to interpret the term $L[g(\bfpsi,\! \bfx_{1\to k},\!i_{1\to k})]$ in the previous claim is as follows. For every $\bfx \in \cX$, we form a new set, which we refer to as $\cX^{\prime}$, as follows: (i) If $\bfx$ does not end with a symbol in $\bfpsi$, add $\bfx$ to $\cX^{\prime}$. Otherwise, (ii) If $\bfx$ ends with a symbol from $\bfpsi$, remove the symbol and add the resulting sequence to $\cX^{\prime}$. 
Then, the quantity $L[g(\bfpsi,\! \bfx_{1\to k},\!i_{1\to k})]$ is the length of the complex sequence that results from concatenating the SCCS of the sequences in $\cX^{\prime}$ with $\bfpsi$.

\Cref{algorithm:SCCS} finds an SCCS of $k$ sequences and its length.
During every iteration of the algorithm, it finds an SCCS of $\bfx_1[1\colon i_1], \bfx_2[1\colon i_2], \ldots, \bfx_k[1\colon i_k]$, following the recursion established in \Cref{claim:SCCS-recursion}, so that there are two cases to consider: 
\begin{enumerate}
    \item There exists $\bfpsi \in \Psi_{q,w}$, such that $U \subseteq \bfpsi$. 
    In this case, an SCCS can be formed by appending the symbol $\bfpsi$ to an SCCS of the sequences $\bfx_1[1\colon i_1^{\prime}], \bfx_2[1\colon i_2^{\prime}], \ldots, \bfx_k[1\colon i_k^{\prime}]$.
    \item The algorithm searches for the complex symbol $\bfpsi$ that minimizes $L[g(\bfpsi,\! \bfx_{1\to k},\!i_{1\to k})]$. 
    At least one index strictly decreases; hence, the recursion terminates.
\end{enumerate}
Finally, for $U$, let $\bfpsi_{U} \in \Psi_{q,w}$ be an arbitrary symbol such that $U \subseteq \bfpsi_{U}$.

\begin{algorithm}
    \caption{Algorithm for finding an $\text{SCCS}(\cX)$.}
    \label{algorithm:SCCS}
    \begin{algorithmic}[1]
        \renewcommand{\algorithmicrequire}{\textbf{Input:}}
        \REQUIRE Alphabet $\Sigma_q$, an integer $w$, and a set of sequences $\cX  = \{ \bfx_1, \bfx_2, \ldots, \bfx_k \} \subseteq\Sigma_{q}^{*}$.
        \renewcommand{\algorithmicensure}{\textbf{Output:}}
        \ENSURE $\text{SCCS}(\cX)$.
        
        \STATE Let $L$ and $S$ be arrays of size $(|\bfx_1|+1) \times \cdots \times (|\bfx_k|+1)$
        \STATE $L[0,0, \ldots, 0] = 0$, $S[0,0, \ldots, 0] = \varepsilon$.
        \FOR {($i_1\coloneqq 0$ to $|\bfx_1|$, $i_2\coloneqq 0$ to $|\bfx_2|$, $\ldots$ , $i_k\coloneqq 0$ to $|\bfx_k|$)}
            \IF {($i_1=0, \ldots, i_k=0$)}
                \STATE \textbf{continue}
		      \ENDIF
            \STATE $U = \emptyset$
            \FOR {($j=1$ to $k$)}
                \STATE $i_j^{\prime}=\max\{0,i_j-1\}$
		  	    \IF {($i_j>0$)}
		  		      \STATE $U = U \cup \{\bfx_j[i_j]\}$
                    \STATE $V = \{\bfpsi\in\Psi_{q,w} \colon U \cap \bfpsi \neq \emptyset\}$
		  	    \ENDIF
		      \ENDFOR
		      \IF {($\left| U \right| \leq w$)}
                \STATE $L[i_1, \ldots, i_k] = L[i_1^{\prime}, \ldots, i_k^{\prime}] + 1$
                \STATE $S[i_1, \ldots, i_k] = S[i_1^{\prime}, \ldots, i_k^{\prime}] \circ \bfpsi_{U}$
            \ELSE
		  	    \STATE \label{alg-step:min-of-all-symbols}
		  	    $L[i_1\!,\! \ldots\!,\! i_k] \!=\! \min_{\bfpsi \in V} \{ L[g(\bfpsi, \bfx_{1\to k},i_{1\to k})]\} \!+\! 1$
		  	    \STATE $\bfpsi = \argmin_{\bfpsi \in V} \{ L[g(\bfpsi, \bfx_{1\to k},i_{1\to k})]\}$
		          \STATE $S[i_1, \ldots, i_k] = S[g(\bfpsi, \bfx_{1\to k},i_{1\to k})] \circ \bfpsi$
		      \ENDIF
        \ENDFOR
        \RETURN $S[|\bfx_1|, |\bfx_2|, \ldots, |\bfx_k|]$
    \end{algorithmic} 
\end{algorithm}

\begin{theorem}\label{theorem:SCCS-algorithm-works}
\Cref{algorithm:SCCS} returns an $\text{SCCS}(\cX)$; in particular
     $L[|\bfx_1|, |\bfx_2|, \ldots, |\bfx_k|]=|\text{SCCS}(\cX)|$.
\end{theorem}
\begin{IEEEproof}
    Immediate from \Cref{claim:SCCS-recursion}.
\end{IEEEproof}

\begin{theorem}\label{theorem:SCCS-algorithm-runtime}
    The runtime complexity of \Cref{algorithm:SCCS} is $O\left(k\cdot\binom{q}{w}\cdot\prod_{j=1}^{k}|\bfx_j|\right)$.
\end{theorem}
\begin{IEEEproof}
    There are $\prod_{j=1}^{k}|\bfx_j|$ iterations. And during each iteration, the worst-case scenario for the algorithm is that it needs to go over all the $\binom{q}{w}$ symbols in $\Psi_{q,w}$, and for each symbol, find which of the $k$ sequences has its next character in the current complex symbol, which yields the runtime complexity.
\end{IEEEproof}

However, if the number of sequences is large, the runtime complexity is no longer polynomial. We address the problem using an approximation algorithm approach next.

\subsection{Approximation Algorithm}
Note that even though \Cref{algorithm:SCCS} runs in polynomial time, it is not always practical, even when $k$ is constant.
In this subsection, we propose an approximation algorithm whose runtime complexity is reduced from $O(n^k)$ to $O(k^{2}n)$.

Specifically, we introduce a straightforward approximation algorithm in \Cref{algorithm:approx-SCCS}, generalizing the algorithms ``M3'' and ``M4'' from~\cite{foulser1992theory, fraser1995subsequences}, originally proposed for the SCS problem.
The output of \Cref{algorithm:approx-SCCS}, denoted by $\bfs$, is a complex supersequence of $k$ input sequences; i.e., each of the $k$ input sequences is a sub-instance of $\bfs$.
To enhance readability, we first introduce some necessary notation. 

Let $A(\cX)=\{\{(\bfy_1)_1, (\bfy_2)_1, \ldots, (\bfy_k)_1\}\}$ denote the multiset containing the first characters of all sequences in a given set $\cX$ consisting of $k$ or fewer sequences (fewer if some sequences are empty).
Further, define $A_w(\cX)$ as the set containing the $w$ most frequent characters from $A(\cX)$; if fewer than $w$ distinct characters appear in $A(\cX)$, the set $A_w(\cX)$ is completed arbitrarily from $\Sigma_q$ to contain exactly $w$ elements.
We say $\cX=\emptyset$ when all sequences within it are empty.

\begin{algorithm}
    \caption{A $k/w$-approximation algorithm for finding an $\text{SCCS}(\cX)$.}
    \label{algorithm:approx-SCCS}
    \begin{algorithmic}[1]
        \renewcommand{\algorithmicrequire}{\textbf{Input:}}
        \REQUIRE A set of sequences $\cX  = \{ \bfx_1, \bfx_2, \ldots, \bfx_k \} \subseteq\Sigma_{q}^{n}$.
        \renewcommand{\algorithmicensure}{\textbf{Output:}}
        \ENSURE $\bfs$, a complex supersequence of $\cX$.
        \STATE $i=1$.
        \WHILE {$(\cX \neq \emptyset)$}
            \STATE Compute $A(\cX)$ and $A_w(\cX)$.
            \STATE Let $\bfs_i=A_w(\cX)$.
            \FOR {$(\bfx_j \in \cX)$}
                \IF {$\left((\bfx_j)_1 \in A_w(\cX)\right)$}
                    \STATE $\bfx_j=(\bfx_j)_{2:|\bfx_j|}$.
                \ENDIF
            \ENDFOR
            \STATE $i = i+1$.
        \ENDWHILE
        \RETURN $\bfs$.
    \end{algorithmic} 
\end{algorithm}

The runtime complexity of \Cref{algorithm:approx-SCCS} is $O(k^{2}n)$, since each iteration has $O(k)$ operations, going over all the sequences' first characters and finding the $w$ most common ones. One needs to iterate at most $kn/w$ times until all $kn$ characters are exhausted. That is $O\left(k\frac{kn}{w}\right)$, but $w$ is a constant; thus, $O(k^{2}n)$.

\begin{claim}
    Let $\text{OPT}=|\text{SCCS}(\cX)|$. Denote $m$ as the size of the resulting sequence from \Cref{algorithm:approx-SCCS}. 
    Then,
    $$m \leq \frac{k}{w} \text{OPT} + 1.$$
\end{claim}
\begin{IEEEproof}
    Clearly, the length $\text{OPT}$ cannot be shorter than $n$. Thus, we show that 
    $$m\leq \lceil \frac{kn}{w} \rceil.$$
    The algorithm begins with $kn$ characters to handle ($k$ sequences of length $n$).
    Since in each iteration of the while-loop we remove at least $w$ characters from the given sequences (assuming $k>w$, since $k$ is very large), except maybe the last iteration, there are at most $\lceil kn/w \rceil$ iterations, and each iteration adds one complex symbol to $\bfs$.
    Therefore, 
    $$m \leq \left\lceil \frac{kn}{w} \right\rceil \leq \frac{k}{w} \text{OPT} + 1.$$
\end{IEEEproof}

We emphasize that this work provides an initial step towards studying approximation algorithms for the scenario of complex sequences.

\section{Efficient Synthesis for Two-Dimensional Strand Arrays with Row Constraints}\label{sec:2d-array}

In this section, we introduce another theoretical synthesis model.
Suppose you need to synthesize strands arranged in an $m \times n$ matrix, where each cell represents a strand.
Per cycle, at most one strand per row may advance; multiple strands in the same column may advance in the same cycle.
That is, in each clock cycle, the synthesis machine generates one symbol according to its predefined (not complex) synthesis sequence $\bfs \in \Sigma_q^{*}$.
In each cycle, at most one strand per row may advance; thus, up to $m$ strands (one per row) can advance simultaneously, possibly in the same or different columns.

In this work, we wish to choose the machine’s synthesis sequence optimally for a fixed set of strands; that is, what is the optimal way to program the synthesis machine and determine which base to synthesize in each cycle?

\begin{example}
Let 
$$\bfa = (0,1), \bfb = (0,2), \bfc = (3,0), \bfd = (1,2),$$
be the $2$-length strands over $\Sigma_4=\{0,1,2,3\}$, that one wishes to synthesize.
Assume that the strands are arranged in a $2 \times 2$ array in the following order:
$$\begin{pmatrix}
\bfa & \bfb \\
\bfc & \bfd
\end{pmatrix}.$$
We provide the following $5$-length shortest synthesis sequence, $\bfs = (0, 1, 3, 0, 2)$.
Then, in the first cycle we synthesize the first element of $\bfa$, but cannot synthesize the first element of $\bfb$, which is also $0$, since they are in the same row.
During the second synthesis cycle, we synthesize both the second element of $\bfa$ and the first element of $\bfd$, which can be done in parallel, since they are in distinct rows.
In the third cycle we synthesize the first element of $\bfc$, in the fourth we add the first element of $\bfb$ and complete $\bfc$.
Finally, in the fifth cycle, we complete both $\bfb$ and $\bfd$ in parallel.
\end{example}

\subsection{Dynamic Programming Algorithm}

For simplicity, consider a $2 \times 2$ array with strands of length $L$ in each cell, and denote them as $\bfa$ and $\bfb$ in the first row, and $\bfc$ and $\bfd$ in the second row.
However, the following dynamic programming algorithm can be easily generalized for the $m \times n$ array with strands of multiple lengths, using the same ideas.

Let $\cS$ be a table indexed by $(i_a,i_b,i_c,i_d) \in \{0,\ldots,L\}^4$.
Define $\cS(i_a, i_b, i_c, i_d)$ as the minimal remaining number of cycles required to synthesize the suffixes $\bfa_{i_a+1:L}$, $\bfb_{i_b+1:L}$, $\bfc_{i_c+1:L}$, and $\bfd_{i_d+1:L}$.
Define
$$\omega_a = \begin{cases}
    \{(i_a+1, i_b, i_c, i_d)\} & \text{if } i_a < L, \\
    \emptyset & \text{otherwise}.
\end{cases}$$
And let $\omega_b$, $\omega_c$, $\omega_d$ be defined similarly to denote all the cases of choosing a symbol that appears only in one strand (per row).
Moreover, let
\begin{align*}
    &\omega_{a,c} = \\
    &\begin{cases}
    \{(i_a + 1, i_b, i_c + 1, i_d)\} & \text{if } i_a,i_c < L, a_{i_a+1}=c_{i_c+1},\\
    \emptyset & \text{otherwise}.
\end{cases}
\end{align*}
And let $\omega_{a,d}$, $\omega_{b,c}$, and $\omega_{b,d}$ be defined similarly to denote all the cases of choosing a symbol that appears in two strands in distinct rows.
Then, we denote 
\begin{align*}
    \Omega_1 &= \omega_a \cup \omega_b \cup \omega_c \cup \omega_d, \\
    \Omega_2 &= \omega_{a,c} \cup \omega_{a,d} \cup \omega_{b,c} \cup \omega_{b,d}.
\end{align*}
Finally, define the following recurrence relation,
\begin{align*}
    &\cS(i_a, i_b, i_c, i_d) = \\
    &\begin{cases}
    0, &\text{if } i_a = i_b = i_c = i_d = L, \\
    1 + \min\{\cS(\omega) \colon \omega\in\Omega_1\cup\Omega_2\}, &\text{otherwise}.
    \end{cases}    
\end{align*}

\begin{algorithm}
    \caption{Computing the length of a shortest synthesis sequence in a $2 \times 2$ array of strands.}
    \label{algorithm:shortest-synth-seq-2d-model}
    \begin{algorithmic}[1]
        \renewcommand{\algorithmicrequire}{\textbf{Input:}}
        \REQUIRE Alphabet $\Sigma_q$, and four sequences of length $L$: $\bfa$, $\bfb$, $\bfc$, and $\bfd$.
        \renewcommand{\algorithmicensure}{\textbf{Output:}}
        \ENSURE The length of a shortest synthesis sequence.
        \STATE $\cS[L][L][L][L] \leftarrow 0$.
        \FOR {($i_a = L$ down to $0$)}
        	\FOR {($i_b = L$ down to $0$)}
                \FOR {($i_c = L$ down to $0$)}
                    \FOR {($i_d = L$ down to $0$)}
                        \IF {($i_a = i_b = i_c = i_d = L$)}
                            \STATE continue
                        \ENDIF
                        \STATE $\Omega \leftarrow \emptyset$.
                        \IF {($i_a < L$)}
                            \STATE $\Omega\text{.append}(\cS[i_a+1][i_b][i_c][i_d])$.
                        \ENDIF
                        \IF {($i_b < L$)}
                            \STATE $\Omega\text{.append}(\cS[i_a][i_b+1][i_c][i_d])$.
                        \ENDIF
                        \IF {($i_c < L$)}
                            \STATE $\Omega\text{.append}(\cS[i_a][i_b][i_c+1][i_d])$.
                        \ENDIF
                        \IF {($i_d < L$)}
                            \STATE $\Omega\text{.append}(\cS[i_a][i_b][i_c][i_d+1])$.
                        \ENDIF
                        \IF {($i_a < L$ and $i_c < L$ and $a_{i_a+1}=c_{i_c+1}$)}
                            \STATE $\Omega\text{.append}(\cS[i_a+1][i_b][i_c+1][i_d])$.
                        \ENDIF
                        \IF {($i_a < L$ and $i_d < L$ and $a_{i_a+1}=d_{i_d+1}$)}
                            \STATE $\Omega\text{.append}(\cS[i_a+1][i_b][i_c][i_d+1])$.
                        \ENDIF
                        \IF {($i_b < L$ and $i_c < L$ and $b_{i_b+1}=c_{i_c+1}$)}
                            \STATE $\Omega\text{.append}(\cS[i_a][i_b+1][i_c+1][i_d])$.
                        \ENDIF
                        \IF {($i_b < L$ and $i_d < L$ and $b_{i_b+1}=d_{i_d+1}$)}
                            \STATE $\Omega\text{.append}(\cS[i_a][i_b+1][i_c][i_d+1])$.
                        \ENDIF
                        \STATE $\cS[i_a][i_b][i_c][i_d] = 1 + \min\{\Omega\}$.
                    \ENDFOR
                \ENDFOR
            \ENDFOR
        \ENDFOR
        \RETURN $\cS[0][0][0][0]$.
    \end{algorithmic} 
\end{algorithm}

\begin{lemma}\label{lemma:S(0)}
    $\cS(0,0,0,0)$ is the length of a shortest synthesis sequence of $\bfa$, $\bfb$, $\bfc$, and $\bfd$.
\end{lemma}
\begin{IEEEproof}
    We prove by induction on $\ell=4L-i_a-i_b-i_c-i_d$.
    For $\ell=0$, we have the base case of nothing to synthesize.
    For the inductive step, we simply check all the valid cases of the next symbol to synthesize and choose one that minimizes the length.
    If $i_a < L$, then $\bfa_{i_a+1}$ can be synthesized in the following synthesis cycle; the event $\omega_{a}$ represents the occurrence of this event, and $\bfa$ is the only sequence that advances during this cycle.
    Similarly, the same is true for the events $\omega_{b}$, $\omega_{c}$, and $\omega_{d}$.
    If both $i_a, i_c < L$ and $\bfa_{i_a+1} = \bfc_{i_c+1}$ can advance in the current cycle, which is valid since they are in distinct rows, the event $\omega_{a,c}$ represents the occurrence of that event.
    Similarly, the same is true for the events $\omega_{a,d}$, $\omega_{b,c}$, and $\omega_{b,d}$.
    Since we are looking for an optimal solution, we need to choose one of these events in $\Omega_1$ or $\Omega_2$ that will determine the next symbol to synthesize, as it is always better to advance one sequence than to perform an idle cycle.
    Next, we use the inductive hypothesis, since all the events on $\Omega_1 \cup \Omega_2$ are either $\ell-1$ or $\ell-2$, and are thus optimal.
    Choosing the minimum of all the events, plus the one synthesis cycle that we add during this step, concludes the proof.
\end{IEEEproof}

\begin{corollary}
    \Cref{algorithm:shortest-synth-seq-2d-model} is a dynamic programming algorithm that solves \Cref{lemma:S(0)} in $O(L^4)$ time and space complexity.
\end{corollary}

\begin{remark}
    By backtracking the array from $\cS(0,0,0,0)$, one can reconstruct a shortest synthesis sequence.
\end{remark}

\subsection{Relation to the SCS Problem}
If $n=1$, the problem was solved in~\cite{Lenz2020} as the SCS of all the strands.
We wish to connect the problem of the optimal synthesis sequence in the array model to an SCS problem.

Let $R_k = \{\bfx_{k,1}, \bfx_{k,2}, \ldots, \bfx_{k,n}\}$ be the sequences in the $k$-th row.
Denote by $\cW_k$ the interleavings of $R_k$, i.e., the set of all merges that preserve the within-strand orders.
Note that each $w\in\cW$ is of length $nL$, the sum of the lengths of all the sequences in $R_k$.
\begin{theorem}
    Let $T^{*}$ be the length of a shortest synthesis sequence for an $m \times n$ array of sequences.
    Then,
    $$T^{*} = \min\limits_{\substack{w_1 \in \cW_1, w_2 \in \cW_2, \\\ldots, w_m \in \cW_m}} \left| SCS(w_1, w_2, \ldots, w_m) \right|$$
\end{theorem}
\begin{IEEEproof}
    Take any synthesis sequence $\bfs$ (the emitted symbols).
    For each row $k$, let $S_k$ be the subsequence of $\bfs$ consisting of the cycles in which row k advanced. 
    Because a row advances at most one strand per cycle and must respect each strand’s order, $S_k \in \cW_k$.
    Also, $\bfs$ is a common supersequence of $S_1, \ldots,S_m$. 
    Hence, $|\bfs| \geq SCS(S_1, \ldots, S_m)$, and
    \begin{align*}
        SCS(S_1, \ldots, S_m) \geq \min\limits_{\substack{w_1 \in \cW_1, w_2 \in \cW_2, \\\ldots, w_m \in \cW_m}} \left| SCS(w_1, \ldots, w_m) \right|.
    \end{align*}
    Since it is true for any $\bfs$, and $T^{*}$ is obtained by minimizing over all $\bfs$, it yields the stated lower bound.

    Conversely, fix any choice $w_k \in \cW_k$ for all $k$, and let $\bfs$ be a shortest common supersequence of $w_1,\ldots ,w_m$.
    Clearly, $\bfs$ is a valid synthesis sequence: on each symbol $r$ of $\bfs$, in each row $k$ advance the row’s pointer in $w_k$ if its next symbol equals $r$, and realize that advance by moving exactly one of that row’s strands that contributes that next symbol in $w_k$.
    This never gets stuck (since $s$ is a supersequence of every $w_k$), and respects the ``at most per row per cycle'' rule.
    Thus, $|\bfs|$ cycles suffice and $T^{*} \leq |s|$ by its definition, including the RHS in the theorem.
    Together with the lower bound, equality holds.
\end{IEEEproof}

\begin{remark}
    For $m=2$ this collapses to the identity $|SCS(x,y)|=|x|+|y|-LCS(x,y)$, i.e.,
    $$T^{*} = 2nL - \max_{w_1 \in \cW_1, w_2 \in \cW_2} \left| LCS(w_1, w_2) \right|.$$
    For $m>2$ there is no simple closed form.
\end{remark}

\section{Conclusion and Future Work}\label{sec:conclusion}
We aimed to find $f(q,w)$ and $f(q,w,\alpha)$, i.e., the maximum information rate for synthesis over $\Psi_{q,w}$, and to identify the sequences that achieve it.
Moreover, we intended to compute the SCCS to find the shortest complex synthesis sequence for a given set of $k$ sequences to synthesize.
We extended Lenz's results~\cite{Lenz2020} for $w=1$ by completely solving $f(q,w)$ and $f(q,w,\alpha)$ and found the complex synthesis sequences that achieve the maximum information rate.
To the best of our knowledge, this is the first analysis of the information rate under the complex synthesis model and also the first analysis of the sub-instance ball as a mathematical structure.

Also, we provided an efficient algorithm for finding the SCCS for a given set of sequences, as well as a simple approximation algorithm for cases where the number of given sequences is large; however, this can be extended to additional approximation algorithms in future work.

Furthermore, for the complex synthesis model, further research can be conducted on synthesis coding and addressing error correction, whether caused by synthesis defects or at any other step during the process of DNA synthesis, storage, or sequencing.

Finally, there is more research to be done on the two-dimensional array synthesis model. One example is the scenario in which the synthesis machine works according to the alternating sequence, the strands are assigned to the cells in the $m \times n$ array, and some questions one can ask are as follows. 
What is the optimal way to program the synthesis machine and determine which base to synthesize in each cycle? How to encode the strands such that the maximum synthesis time is minimized?

\section*{Acknowledgments}
The research was funded by the European Union (ERC, DNAStorage, 101045114, and EIC, DiDAX 101115134). Views and opinions expressed are, however, those of the authors only and do not necessarily reflect those of the European Union or the European Research Council Executive Agency. Neither the European Union nor the granting authority can be held responsible for them.
This work was also supported in part by NSF Grant CCF2212437.


\bibliographystyle{IEEEtran}
\bibliography{bibli}
\end{document}